\documentclass[reprint, amsmath, amssymb, aps, showkeys]{revtex4-2}

\usepackage{graphicx}
\usepackage[colorlinks=true, allcolors=blue]{hyperref}
\usepackage{multirow}

\newcommand{\lt}{\left}
\newcommand{\rt}{\right}

\begin{document}

\preprint{APS/diffusion}

\title{Diffusion transitions in a 2D periodic lattice}

\author{Matheus J. Lazarotto}
 \email{matheus.lazarotto@usp.br}
\author{Iber\^{e} L. Caldas}
 \affiliation{Instituto de F\'{i}sica, Universidade de S\~{a}o Paulo, Rua do Mat\~{a}o 1371, S\~{a}o Paulo 05508-090, Brazil}
\author{Yves Elskens}
 \affiliation{Aix-Marseille Universit\'{e}, CNRS, UMR 7345 PIIM, F-13397 Marseille, France}

\date{\today}


\begin{abstract}
    Spatial diffusion of particles in periodic potential models has provided a good framework for 
    studying the role of chaos in global properties of classical systems.
    Here a bidimensional ``soft'' billiard, classically modeled from an optical lattice Hamiltonian 
    system, is used to study diffusion transitions under variation of the control parameters. 
    Sudden transitions between normal and ballistic regimes are found and characterized by inspection 
    of topological changes in phase-space.
    Transitions correlated with increases in global stability area are shown to occur for energy levels 
    where local maxima points become accessible, deviating trajectories approaching them. 
    These instabilities promote a slowing down of the dynamics and an island myriad bifurcation 
    phenomenon, along with the suppression of long flights within the lattice.
    Other diffusion regime variations occurring within small intervals of control parameters are
    shown to be related to the emergence of a set of orbits with long flights, thus altering 
    the total average displacement for long integration times but without global changes 
    in phase-space.
\end{abstract}

\maketitle

\section{Introduction}\label{sec:introduction}
From a wide range of experimental contexts, periodic potentials have been used to model diverse physical 
systems, from optical lattices, where ultra-cold gases or Bose-Einstein condensates are confined by a 
set of laser beams \cite{Bloch1, Bloch2, Hemmerich, Monteiro}, to ionic particles submitted to guided 
waves in $\textbf{E}\times\textbf{B}$ fields and plasma physics \cite{Kleva, Horton}, to wave propagation 
in photonic crystals \cite{Yu}, and to xenon atoms diffusion over platinum surfaces \cite{Sholl}.

These systems have also been used as models for the quantum-classical correspondence 
of chaos for many-body systems. As shown by Thommen \cite{Thommen}, particles in a quantum tilted 
lattice can present a behavior similar to classical chaos.
Prants \cite{Prants_2016, Prants_2017} proposes scenarios where the effect of chaos in particles 
displacement could be observed experimentally in a periodic potential with semiclassical effects 
when the field interacts with two-level atoms.

Particularly within the theoretical perspective of classical dynamical systems, periodic potentials commonly 
appear in Sinai billiards and hard wall scatterers models for Lorentz gases \cite{Zaslavsky_2, Bagchi, Machta}. 
Analogous Hamiltonian models with smooth periodic potentials, the so-called ``soft'' billiards, were 
also considered, providing a good framework to study chaotic dynamics and control theory 
\cite{Kroetz, Kaplan, Reichl}. Such a feature is particularly relevant for the transport of particles 
through lattices, since anomalous diffusion regimes can occur without the application of random forces, 
purely due to inner instabilities in the system motion 
\cite{Kleva, Argonov_1, Argonov_2, Prants_2019, Zaslavsky_1, Zaslavsky_3, Zaslavsky_4, Chaikovsky}.

Recent works on periodic potentials as those on soft billiards based on optical lattices display 
their mixed Hamiltonian phase-space structure and some scenarios for different parameters. We particularly
mention Horsley \textit{et al.}, for a study on the same Hamiltonian model used here \cite{Horsley} 
(introduced in section \ref{sec:model}), although no considerations on diffusion and its dependency 
on the system parameters were included.
Regarding classical aspects on quantum models, Porter \textit{et al.} studied bidimensional (2D) 
hamiltonian lattices for different topologies, such as a honeycomb lattice \cite{Porter1}, including 
a quantum treatment for the energy bands and the effect of symmetry breaking \cite{Porter2}. In 
\cite{Prants_2019}, Prants studies a Hamiltonian model analogous to the one considered here but 
modelling a dipole-field interaction by including into the coherent dynamics the interaction of 
internal atomic degrees of freedom with translational ones. 

Regarding diffusion itself, Zaslavsky performed some of the prominent works for continuous flows 
and Hamiltonian models considering the variation of control parameters \cite{Zaslavsky_1, Zaslavsky_2, Zaslavsky_4}. 
Particularly in \cite{Zaslavsky_4}, for a periodic Q-model, were shown the existence of long flights 
within the chaotic motion, the consequent anomalous diffusion regime, the tail thickening effect 
in the power-law distribution, and the occurrence of sudden transitions in diffusion rate as a function 
of a control parameter. 
Similarly, Argonov and Prants \cite{Argonov_1, Argonov_2, Argonov_3} showed the fractal structure of 
escape time basins and the effect of chaos in different diffusion regimes due to L\'{e}vy flights in 
a fully coherent semiclassical model of a 1D lattice.
Kleva \cite{Kleva} considered in detail the structure of chaos 
close to movable separatrix curves of a periodic potential displaying these same L\'{e}vy flights. 
The connection between these flights and anomalous regimes has been shown for a wide variety of models 
\cite{Zaslavsky_2, Mandal}.

From this diffusion background, the present work aims to a detailed description of the 
dynamics behind the transport of particles in a lattice Hamiltonian system. A bidimensional periodic 
classical potential, based on an optical lattice system, is used to numerically study transitions in the 
spatial diffusion as the main control parameters change, namely the total particle energy and the 
coupling between the waves that form the lattice. Generally, for either varying energy or coupling, 
the disruption of stability islands and further bifurcation of periodic orbits (PO), both stable (SPO) 
and unstable (UPO), results in a non-trivial dynamics with mixed phase-space. Due to such bifurcations, 
the system's spatial diffusion regime, measured by a long time exponent rate, undergoes sudden 
variations between normal and ballistic regimes or even peaks of intermediate rate with long transient 
times. The lack of pattern or predictability of such transitions thus motivates an inspection of the 
topological changes within phase-space. In order to conduct this inspection as the control parameters 
vary, a series of numerical techniques are applied, namely, the measurement of the 
ratio of chaotic/regular area within phase-space, the automated search for POs and identification 
of their stability and period, the manifolds from UPOs, and the escape time (or range) basins.

From the general finding of non-trivial changes in diffusion, a particular transition was studied 
in more detail. An increase (decrease) of stability (chaotic) area was found to 
correlate with a sudden diffusion change from ballistic to normal regime as 
energy increases. 
We verified that such behavior occurs when particles energy is enough to reach local instability 
points (potential local maxima), promoting a slowing down of dynamics and giving rise to new 
regular structures amidst a web-like manifold pattern in phase-space. This myriad 
of islands soon vanishes as energy increases and is replaced with a global chaotic sea, determining 
a more restrained chaotic dynamics with suppressed long flights through the lattice. 
For diffusion variations uncorrelated with changes in chaotic/stable areas, increments in the 
particles average displacement are found to be caused by the onset of a set of long flight orbits. 
These transitions exist for a short energy interval and represent a small part of the total stable 
area but still alter the average displacement calculation as a transient behavior, perceptible for long 
times ($t \propto 10^3$).

In this paper, sections \ref{sec:model} and \ref{sec:diffusion-method} respectively present the 
lattice model used in this study and how the transport of particles is evaluated.
The main findings are discussed along section \ref{sec:results}, starting by the aforementioned 
diffusion transitions (section \ref{sec:results-diffusion-exponent}) and followed by an initial 
analysis of the onset of chaos and transport in the system (section \ref{sec:results-diffusion-onset}). 
A particular transition is detailed in sections \ref{sec:results-local-maxima-transition} 
and \ref{sec:results-periodic-orbits} by showing the phase-space changes and POs bifurcations. 
For completeness, in section \ref{sec:results-long-transient} we also present results for 
diffusion variations not related to global modifications in the dynamics. 
Final remarks are given in section \ref{sec:conclusions}. An appendix section is provided to enhance 
the discussion on integration error, statistical convergence, further PO analysis and extra results. 
Each topic is mentioned through the text whenever it may be of interest.

\section{Model}\label{sec:model}

The bidimensional periodic lattice model studied in this work stems from a classical Hamiltonian 
motivated by the trapping of a single neutral particle in the superposition of standing laser waves 
\cite{Bloch1}. 
Experimentally, this setup is used as an optical lattice to trap and control cold atom gases or 
Bose-Einstein condensates, although its treatment is made quantically. In such an arrangement, an 
electrical field $\textbf{E}$ from a standing-wave, obtained by the superposition of multiple 
counter-propagating monochromatic waves, induces a parallel dipole $\textbf{d}$ into a neutral particle. 
The re-interaction between field and dipole yields the potential
\begin{equation}\label{eq:potential-general}
    V_{\textrm{dip}}(\textbf{r}) = -\textbf{d}\cdot\textbf{E}(\textbf{r}) = -\sum_i \lt(\textbf{d}\cdot\hat{e}_i\rt) E_0^i \cos(\textbf{k}_i\cdot\textbf{r} + \theta_i),
\end{equation}
with $\textbf{k}_i$ as the wave vector and $\theta_i$ the phase of each individual standing wave; 
$\hat{e}_i$ and $E_0^i$ are the respective polarization direction and amplitude of the waves, and 
$\textbf{r}$ the particle spatial position. The induced dipole given by $\textbf{d}=\gamma(\omega)\textbf{E}$ 
holds for the case where the incident radiation frequency $\omega$ does not resonate with the particle 
inner energy levels, avoiding absorption and further re-emission of photons and thus ensuring a 
linear polarization by a polarizability factor $\gamma(\omega)$. This aspect 
allows for an approximation of spatial trajectories as classical. Alternatively, Prants \cite{Prants_2016,Prants_2017, Prants_2019} 
considered a semiclassical effect of particle-field interaction by allowing photons absorption and 
emission for particles with two inner energy levels.

The generic form of potential (\ref{eq:potential-general}) allows the construction of different lattices 
by the combination of multiple waves and frequencies. For a 2D lattice, at least two linearly independent 
wave vectors are selected, constraining the particle movement along two cartesian axes, although an 
extra wave propagating in the same $(\textbf{k}_1, \textbf{k}_2)$ plane can be placed in order to 
achieve different topologies, as done by Porter \textit{et al.} for a honeycomb lattice \cite{Porter2}. 
Indeed, even non-harmonic fields can be used for the lattice construction, particularly if one aims to 
consider symmetry breaking effects \cite{Porter1}.

Here, two orthogonal waves within the $x\!-\!y$ plane are taken with equal amplitude and wavelength 
($\textbf{k}_x = k\hat{x}; \textbf{k}_y = k\hat{y}$), yielding the periodic potential for the lattice
\begin{equation}\label{eq:potential-2d}
\begin{split}
    V(x',y') = U' \lt(\cos^2(k x') + \cos^2(k y') +\rt. \\
    \hspace{1.5cm}\lt. 2 \alpha \cos(k x') \cos(k y')\rt),
\end{split}
\end{equation}
with 
\begin{equation}\label{eq:parameters}
    U' = \gamma(\omega)E_0^2 > 0 \quad\textrm{and}\quad \alpha = \lt(\hat{e}_x \cdot \hat{e}_y\rt) \cos(\theta_x - \theta_y).
\end{equation}

\begin{figure}[ht]
    \centering
    \includegraphics[width=0.40\textwidth]{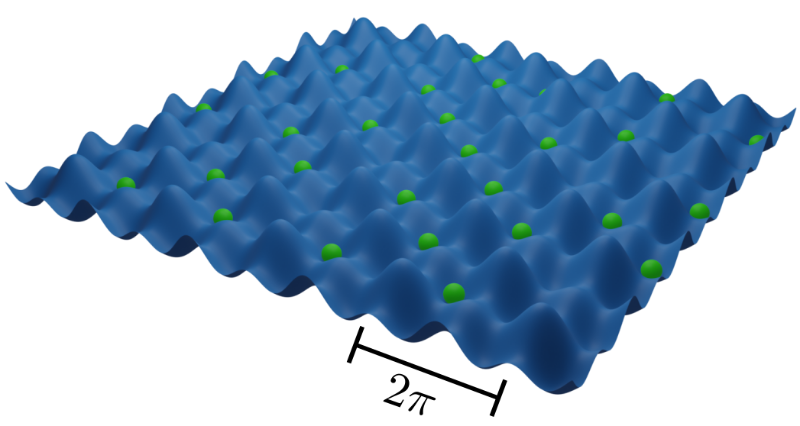}\vspace{-0.3cm}
    \caption{3D representation of particles (in green) over the lattice potential surface $V(x,y)$ 
             displaying its periodic structure.}
    \label{fig:lattice-3d}
\end{figure}

Therewith, for a single particle the lattice Hamiltonian can be written as
\begin{equation}\label{eq:hamiltonian}
\begin{split}
    H = p_x^2 + p_y^2 + U\!\lt(\cos^2(x) + \cos^2(y) + \rt.\\
                               \lt.2 \alpha \cos(x) \cos(y)\rt),
\end{split}
\end{equation}
with space units scaled to ($x=k x'; y=k y'$) and Hamiltonian to $H = 2m H'$, so that 
the energy scale is $U=2m U'$. The classical dynamics of a particle is thus described by 
the surface potential shown in figure \ref{fig:lattice-3d}, where a particle can be trapped 
in the field wells for energies lower than those of the saddle point between pits or otherwise 
diffuse through the lattice cells above this threshold.

The energy scale $U$ is of no relevance in the classical regime, in the sense that it does not alter 
the topology of solutions whatsoever, and can be set to $1$ by rescaling time. In the quantum regime 
on the other hand, this energy scale relates to the accessible eigenstates and thus has further 
relevance. For this study, we fix $U=20$ following Horsley \textit{et al.} \cite{Horsley} 
since it corresponds to a feasible value obtainable in experiments. 
In \cite{Prants_2019}, Prants considered a similar Hamiltonian to (\ref{eq:hamiltonian}) but 
taking into account the coupling of internal and translational degrees of freedom of the atom 
via a coherent semiclassical model for laser frequencies close to optical resonance.

As can be seen from potential (\ref{eq:potential-2d}), the coupling parameter $\alpha$ stands for the 
product of the fields polarizations and phase difference, acting as a perturbation to the integrable 
Hamiltonian of two pendula potentials along $x$ and $y$ (with period $\pi$), coupling them for any 
$\alpha\neq 0$. Although $\alpha$ values may vary within $[-1,1]$, one can notice that it is only 
required to consider solutions for $[0,1]$, since the negative counterpart is equivalent to a spatial 
translation by $\pi$ in one of the cartesian directions, thus not altering solutions properties.

\begin{figure}[ht]
    \centering
    \includegraphics[width=0.5\textwidth]{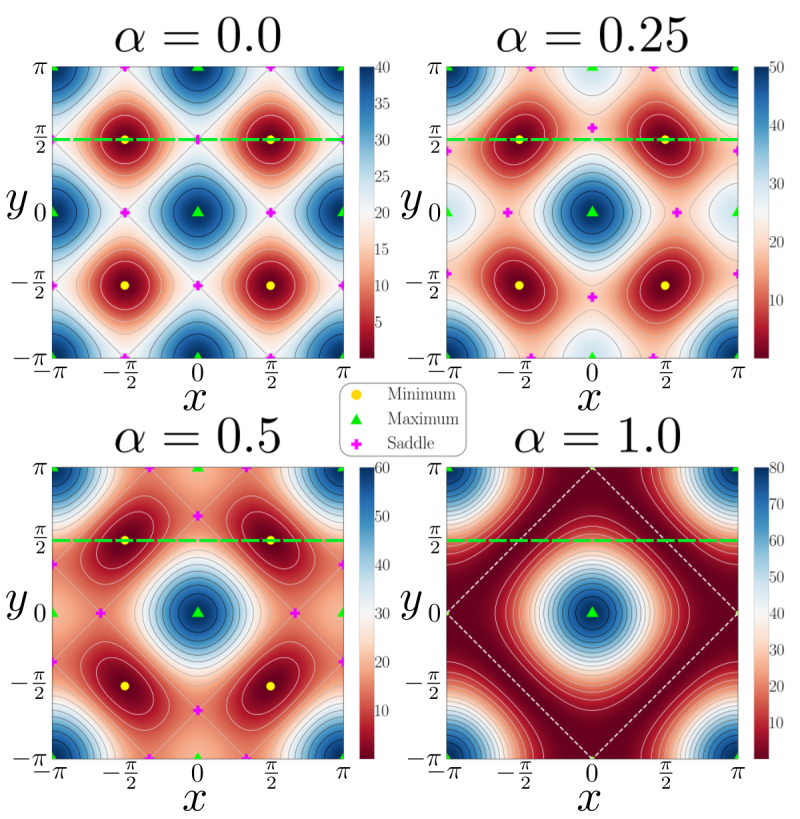}
    \caption{Color plot of the top view of potential $V(x,y)$ minimal unit cell for different 
             values of the coupling $\alpha$. 
             The PSS (Poincar\'{e} Surface Section) used for phase-space display is shown as the 
             horizontal dashed green line at $y=\frac{\pi}{2}$.}
    \label{fig:potential}
\end{figure}

Figure \ref{fig:potential} shows how the periodic potential structure changes as $\alpha$ increases 
from the separable case ($\alpha=0$) to the maximum superposition amplitude ($\alpha=1$). As the 
saddle points move towards the local maxima, they finally merge when $\alpha=1$, forming two trenches 
with degenerate minima on the lines $y + x \equiv \pi \mod(2\pi)$ and $y - x \equiv \pi \mod (2\pi)$ 
(see table \ref{tab:equilibrium}). For increasing $\alpha$, the path between lattice pits for 
particles to diffuse through lattice cells gets wider and wider, since the diffusion's energy 
threshold at the saddle points is given by $V^{\textrm{saddle}}(\alpha)=U(1-\alpha^2)$. 

\begin{table}[ht]
    \centering
    \begin{tabular}{l|c|c} \hline\hline 
    Equilibrium points          & $(x^*, y^*)$                                  & $V(x^*, y^*)$                    \\ \hline
\multirow{4}{*}{Minima}         & $\lt(\frac{\pi}{2},   \frac{\pi}{2}\rt)$      & \multirow{4}{*}{$0$}             \\
                                & $\lt(-\frac{\pi}{2}, -\frac{\pi}{2}\rt)$      &                                  \\
                                & $\lt(\frac{\pi}{2}, -\frac{\pi}{2}\rt)$       &                                  \\                                
                                & $\lt(-\frac{\pi}{2}, \frac{\pi}{2}\rt)$       &                                  \\ \hline
\multirow{2}{*}{Maxima (global)}& $(0,0)$                                       & \multirow{2}{*}{$2U(1+\alpha)$}  \\ 
                                & $(\pi, \pi)$                                  &                                  \\ \hline 
\multirow{2}{*}{Maxima (local)} & $(\pi, 0)$                                    & \multirow{2}{*}{$2U(1-\alpha)$}  \\
                                & $(0, \pi)$                                    &                                  \\ \hline
\multirow{4}{*}{Saddle}         & $(0, \pm\cos^{-1}(-\alpha))$                  & \multirow{4}{*}{$U(1-\alpha^2)$} \\ 
                                & $(\pm\cos^{-1}(-\alpha), 0)$                  &                                  \\ 
                                & $(\pi, \pm\cos^{-1}(\alpha))$                 &                                  \\
                                & $(\pm\cos^{-1}(\alpha), \pi)$                 &                                  \\ \hline\hline
    \end{tabular}
    \caption{Equilibrium points position $(x^*, y^*)$ and energy value $V(x^*, y^*)$ within a unit 
             cell of the periodic lattice; positions are taken with modulo $2\pi$. At $\alpha=1$, 
             saddle points merge with local maxima and form minimum trench lines given by 
             $\cos(y_\textrm{t}(x)) = -\cos(x)$.}
    \label{tab:equilibrium}
\end{table}

For the results shown along this work, all phase-space portraits will be made over the same Poincar\'{e} 
surface section (PSS) -- highlighted by green lines in figure \ref{fig:potential}, namely the one defined 
as the oriented surface along two of the lattice pits, \textit{i.e.},
\begin{equation}\label{eq:pss}
    \Sigma = \lt\{\lt(x, y, p_x, p_y\rt) \in \mathbb{R}^4 : \lt(y = \frac{\pi}{2}; \; p_y > 0\rt)\rt\}.
\end{equation}

Since Hamiltonian (\ref{eq:hamiltonian}) is autonomous, energy ($E=H$) is an immediate constant of 
motion, constraining trajectories in a three-dimensional surface, which can thus be pictured by a 2D 
section. The oriented surface $\Sigma$ is particularly convenient for this potential since it intersects all possible 
solutions within a single unit cell pit, except for the UPOs along the stable direction of the 
saddle equilibrium points located at $y=0$ and $y=\pi$. Indeed, bounded solutions around the minima 
points with $y<0$ will occur, but nonetheless the $\frac{\pi}{2}$ rotation invariance implies that 
their symmetrical counterpart solutions will intersect $\Sigma$ at $y=\frac{\pi}{2}$. 
The potential symmetry allows us to consider the motion in a restricted 
phase-space if one modulates trajectories with periodic boundary conditions ($x,y \in 
[-\pi,\pi)$), but it also allows for the scattering or diffusion of particles if one lets spatial 
variables run freely.

The PSS $\Sigma$ will also be used as reference for representing periodic orbits. Along any PSS, POs 
appear as fixed points with discrete period, \textit{i.e.}, an $n$-periodic orbit will cross the PSS 
$n$ times until it return to its initial point. However, the number of these crossings depends on 
the orientation of the surface and how it intercepts the orbit path. 
Therefore, along the text we refer to discrete periods (labeled as $T$) only defined relative to 
$\Sigma$ (\ref{eq:pss}). Moreover, we differentiate the discrete period from the real-valued dynamical 
period (labeled as $\tau$), which is simply the total time elapsed for the orbit to returns to its 
initial point.

\section{Diffusion calculation}\label{sec:diffusion-method}
The quantitative characterization of the diffusion regime of an ensemble of particles through the 
lattice was straightly achieved from the asymptotic power law \cite{Zaslavsky_4}
\begin{equation}\label{eq:diffusion-power-law}
    \langle R^2 \rangle (t) \propto t^{\mu},
\end{equation}
where $R=\sqrt{(x - x_0)^2 + (y - y_0)^2}$ is the spatial displacement of a particle from its initial 
position, with the average taken over all particles. The exponent $\mu$ thus indicates the diffusion 
rate, being normal ($\mu=1$), ballistic ($\mu=2$) or anomalous for different results within $0 < 
\mu < 2$. 
One can immediately assert from energy conservation and the boundedness of the potential (\ref{eq:hamiltonian}) 
that super-diffusivity ($\mu>2$) cannot occur in the system.

To numerically compute $\mu$ for a given pair of parameters ($\alpha$, $E$), an ensemble of $N$ 
random initial points covering the 3D energy shell in phase-space is evolved for long times 
($t=2000.0$). From the time series data for $\langle R^2 \rangle (t)$, one can retrieve the rate 
exponent $\mu$ by fitting it with the power-law (\ref{eq:diffusion-power-law}). Since the power-law 
is expected to hold true only asymptotically to long times, the data fitting is performed over the 
last 30\% of time interval data. 
The random initial points are sampled as a Monte-Carlo procedure by randomly generating a position 
$(x_0, y_0)$ within the constraint $V(x_0,y_0,\alpha) < E$ and a momentum vector $\vec{p}=(p_x,p_y)$ 
with random direction given by an angle uniformly distributed within $[0,2\pi)$ and modulus defined 
by the conservation of energy (\ref{eq:hamiltonian}).

Besides errors of statistical order, the long-time integration required raises the issue of numerical 
error in the solutions. A Runge-Kutta-Cash-Karp (RKCK) \cite{Cash-Karp} method is used, and therefore it 
does not conserve the symplectic 2-form (even for regular solutions). However, energy deviations  
did not exceed the order of $10^{-9}$, implying that the solutions obtained are very well bounded within 
the same energy hyper-surface, even though individual trajectories present small divergence from the 
real solution. This divergence is surmounted in this case since only the average over a 
uniformly filled phase-space is required, given that no stickiness was found to be relevant for the 
system's dynamics. 
Nevertheless, a direct comparison to a symplectic method, developed by 
M. Tao \cite{Tao}, was made and is inserted in appendix \ref{sec:appendix:symplectic}, 
showing that the lack of symplecticity of the RKCK method does not impact the results 
obtained here.

\section{Results}\label{sec:results}
\subsection{Diffusion exponent}\label{sec:results-diffusion-exponent}
\begin{figure}[ht]
    \includegraphics[width=0.5\textwidth]{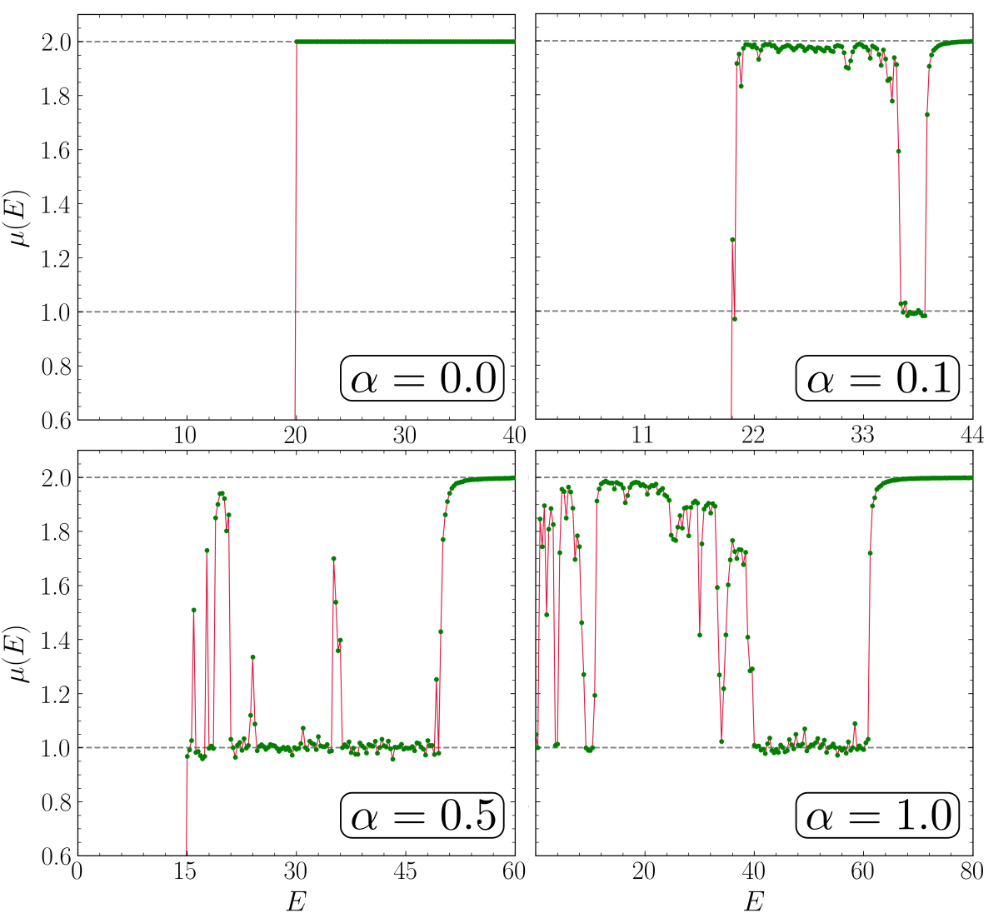}
    \caption{Diffusion exponent $\mu(E)$ for different values of the coupling $\alpha$. Any non-integrable 
             scenario ($\alpha \neq 0$) presents sudden transitions between diffusion regimes.
             Energy scales range $(E_\textrm{min} \to E_\textrm{max}) = (0 \to 2U(1+\alpha))$.}
    \label{fig:diffusion_varE}
\end{figure}

The study of the transport of particles through the lattice is made by the procedure described 
in section \ref{sec:diffusion-method} while varying the main control parameters, \textit{i.e.}, the 
total energy of the particle ($E$), and the coupling between radiation waves ($\alpha$). 
Figure \ref{fig:diffusion_varE} shows the profile found for different $\alpha$ values and varying 
energy. When the system is integrable ($\alpha=0$), it is spatially separable and, for energies 
allowing for diffusion, its behavior is completely ballistic, since this case is similar to a pendulum 
with rotation energy, moving unimpeded through the lattice. When integrability is broken ($\alpha \neq 0$), 
generally the diffusion exponent $\mu(E)$ undergoes a series of short and sudden transitions, 
with changes from normal ($\mu=1$) to free ($\mu=2$) regimes occurring abruptly and intermediate 
variations to intermediary regimes ($1 < \mu < 2$), mostly seen as sharp peaks or valleys.

As exposed in section \ref{sec:model}, the coupling $\alpha$ relates to the broadening of spatial 
channels through which a particle travels to neighbour lattice pits (stable minima points). Both 
the saddle points between the pits and local maxima decrease in energy, thus allowing a wider space 
for flights. Thus, one could hope to primarily find a continuous transition for the transport exponent 
as a function of both $\alpha$ and $E$. However, inner instabilities from chaotic trajectories and 
a mixed phase-space, with the emergence or destruction of Kolmogorov-Arnol'd-Moser (KAM) islands 
(PO bifurcations), imply a more intricate profile for the particles diffusion and its exponent $\mu$. 
Indeed, when conducting the equivalent diffusion calculation for fixed energies and $\alpha$ as the 
free parameter, similar behavior is found (results shown in appendix \ref{sec:appendix:diffusion-var-alpha}), 
given that changes in dynamics occur in a similar way, namely the bifurcation of stability islands 
or UPOs, for the variation of both control parameters.

For energies close to global maximum points ($E \approx V_\textrm{g-max} = 2U(1 + \alpha)$), a common 
plateau at ballistic regime occurs for all values of $\alpha$. At this energy level, the particle dynamics 
covers a wider space region with large momenta, where instabilities from the coupling can be seen as small 
perturbations. Therefore, the particle movement is that of a point moving rapidly through the lattice 
with smaller deviations, thus yielding a ballistic diffusion regime. Furthermore, this is also verified 
by an emergent dominance of large chains of stability islands in phase-space.

Given the statistical reliance on the calculation for $\mu(E)$, an immediate 
concern with its convergence and error is raised, particularly regarding the total integration 
time $t$ and ensemble size $N$. Details on this error analysis are given in appendix \ref{sec:appendix:statistical-convergence}, 
while here we summarize that the transitions undergone by $\mu(E)$ are found for either 
time and ensemble size of different magnitudes, whereas absolute deviations in their values do not 
exceed 15\%. However, transitions composed of single thin peaks or intermediary rate values 
($1<\mu<2$) may correspond to long transient behavior due to small stability islands composed of long 
flights, slowly converging towards ballistic rate $\mu\to2$, although only manifested for longer times; 
a further discussion is made in section \ref{sec:results-long-transient}.

\begin{figure}[ht]
    \includegraphics[width=0.38\textwidth]{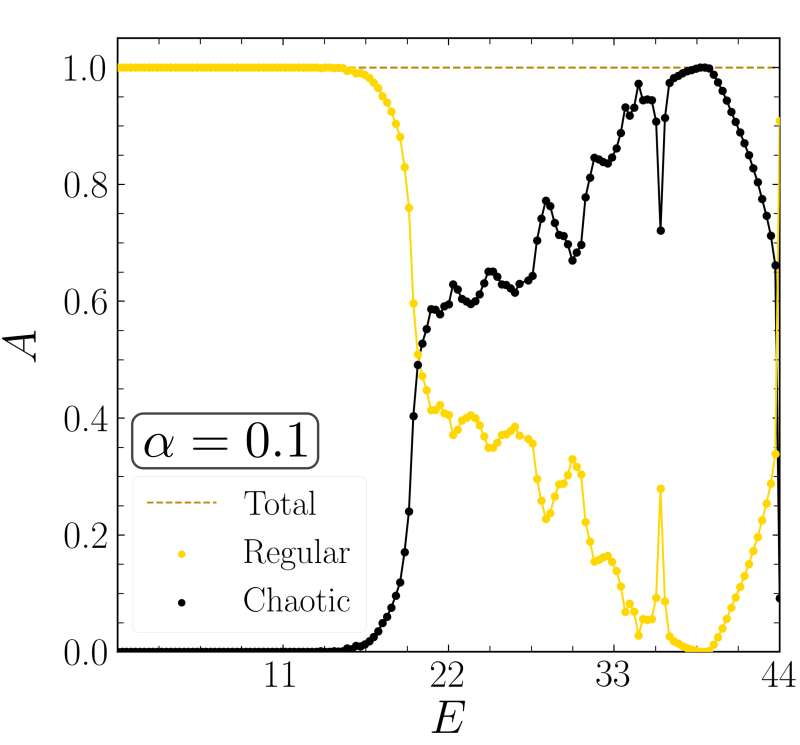}
    \includegraphics[width=0.38\textwidth]{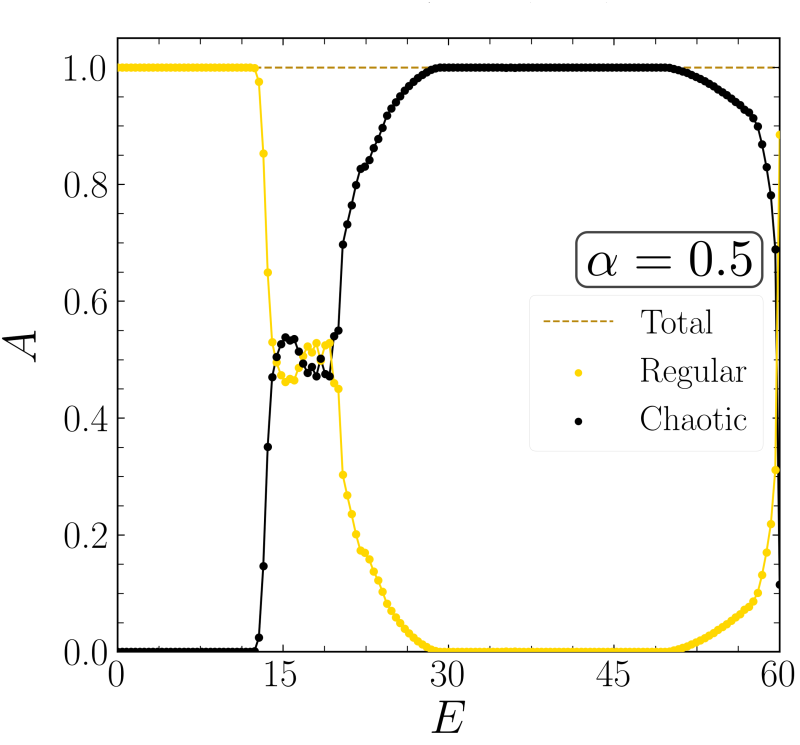}
    \includegraphics[width=0.38\textwidth]{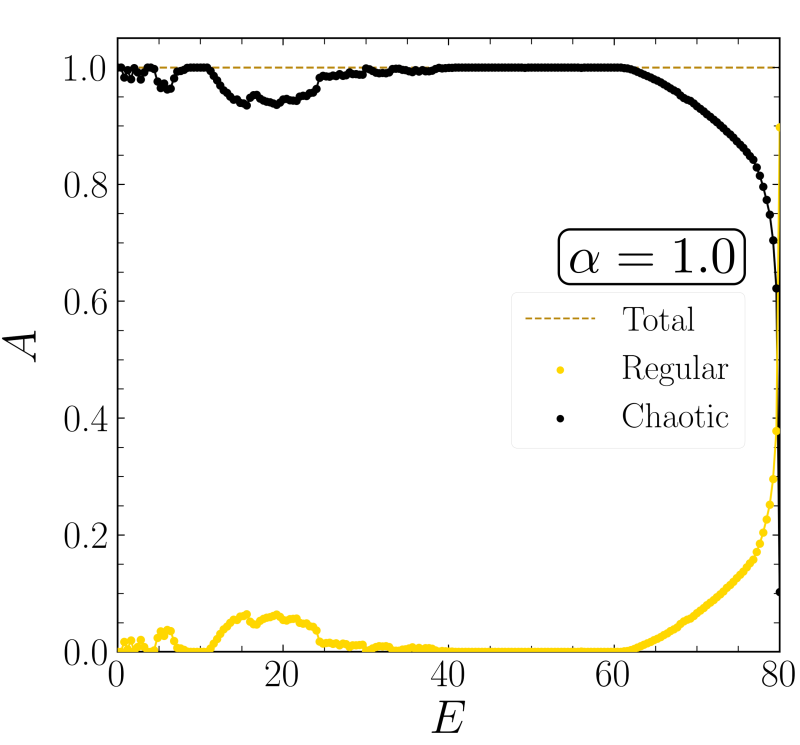}\vspace{-0.4cm}
    \caption{Normalized chaotic/regular area $A$ over the PSS $\Sigma$ for varying energy and 
             different fixed coupling values calculated via SALI algorithm.}
    \label{fig:chaotic-area}
\end{figure}

One may immediately expect that the transitions shown in figure \ref{fig:diffusion_varE} depend 
entirely on the mixed dynamics of the system's solutions. Therefore, we firstly inquired whether 
these transitions correlate to the appearance or destruction of stability islands. For this purpose, 
the chaotic area was measured over the PSS $\Sigma$ (\ref{eq:pss}) via Smaller Alignment Index (SALI) 
method \cite{Skokos_book, Skokos_paper}. Briefly putting it, the algorithm considers two deviation 
vectors $\lt(\hat{\omega}_1, \hat{\omega}_2\rt)$ which are integrated 
by the first order variation of motion equations along with a reference orbit (their norms being kept 
constant). These deviation vectors thus behave differently for each kind of orbit. In case the reference 
orbit is chaotic, they align or anti-align to each other since they orient themselves towards the unstable 
manifold direction, whereas for a regular orbit they are kept at finite angle while only orienting 
themselves towards the tangent plane of the stability torus in which the orbit is contained. 
It is possible that for regular orbits the tangent vectors still align/anti-align 
due to shear between close torus layers, however this may occur for times much longer than the one for 
alignment in chaotic orbits.

From this principle, the defined index
\begin{equation}\label{eq:sali}
    \textrm{SALI}(t) = \textrm{min}\lt(\lVert\hat{\omega}_1 + \hat{\omega}_2\rVert, 
                                       \lVert\hat{\omega}_1 - \hat{\omega}_2\rVert\rt)
\end{equation}
provides a quantitative way to discriminate the orbit's nature, given that aligned vectors imply 
SALI($t$) $\to0$ while regular ones keep SALI at constant non-zero values (assuming non-parallel 
initial vectors). The algorithm does not rely on any particularity of the present model, thus being 
applicable in general to Hamiltonian systems or symplectic discrete maps.

Using this index to differentiate between regular and chaotic orbits, one can mesh the surface 
$\Sigma$ and sum over the sub-areas from each initial condition assigned to each grid cell. 
Equivalently, the same procedure could be made for a 4D grid over the whole phase-space. However,  
the selected PSS intersects all possible orbits within a unit cell, except 
the UPOs along the saddle equilibrium points at $y=0$ and $y=\pi$, thus ensuring that the section 
provides a good portrait of the chaos/stability ratio of the whole phase-space. 
Figure \ref{fig:chaotic-area} shows the area portions of orbits along $\Sigma$ normalized by the 
total area $A_\textrm{tot} = A_{\textrm{chaos}} + A_{\textrm{regular}}$ for the same 
coupling values displayed in figure \ref{fig:diffusion_varE}.

Figure \ref{fig:chaotic-area} shows a correlation between a main transition for ($\alpha=0.1; E=36$), 
where a sudden increase of stable area is found to occur, and the transition in diffusion exponent 
$\mu(E)$ shown in figure \ref{fig:diffusion_varE}, where a free transport plateau drops to normal 
regime. On the other hand, a series of peaks in diffusion rate, either from increasing or decreasing 
$\mu(E)$, do not correlate with any pronounced changes in chaotic/stable area. This is particularly 
seen along the intervals at ($\alpha=0.5; E\in[30,50]$) and ($\alpha=1.0; E\in[30,60]$) where the 
system is dominantly chaotic along the whole energy interval although peaks of diffusion transition 
are seen for $\mu(E)$.
Even though no direct implication requires the diffusion exponent to correlate directly with the chaotic 
area, it is compelling to check whether the transitions found are related to the emergence of chaos or 
inner changes in its domain, such as the appearance of L\'{e}vy flights, which may not alter the area but 
change chaos properties.
Given these two scenarios for transport variation, we aim to inspect the 
phase-space for energy values around these transitions and characterize the dynamical changes 
occurring along with it.

\subsection{Diffusion onset}\label{sec:results-diffusion-onset}

\begin{figure}[b]
    \centering
    \includegraphics[width=0.5\textwidth]{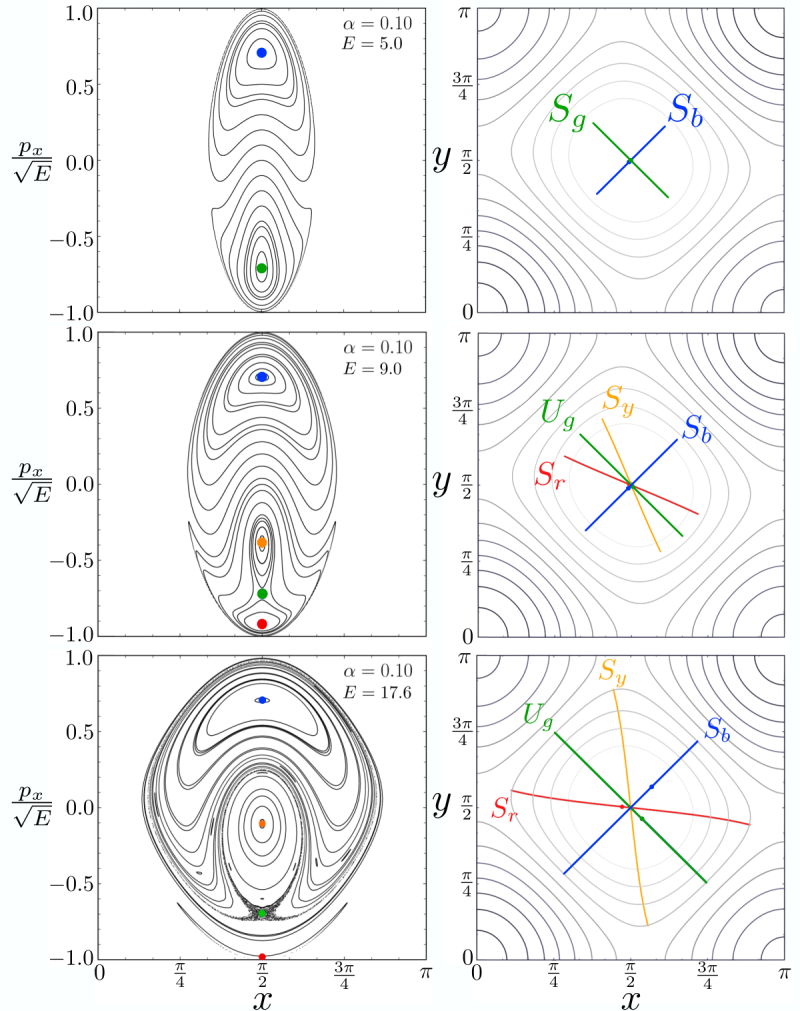}
    \caption{PO bifurcation process for increasing energy seen from the PSS $\Sigma$ (left column), with 
             fixed points drawn as colored circles, along with its correspondent spatial 
             trajectories in the right column. Each PO is named after its stability -- stable ($S$) or unstable ($U$) 
             -- and an index, related to its color. In the portraits, top row has $E=5$, center 
             row $E=9$ and bottom row $E=17.6$; in all of them $\alpha=0.1$.}
    \label{fig:upo-bifurcation}
\end{figure}

\begin{figure}[b]
    \centering
    \includegraphics[width=0.5\textwidth]{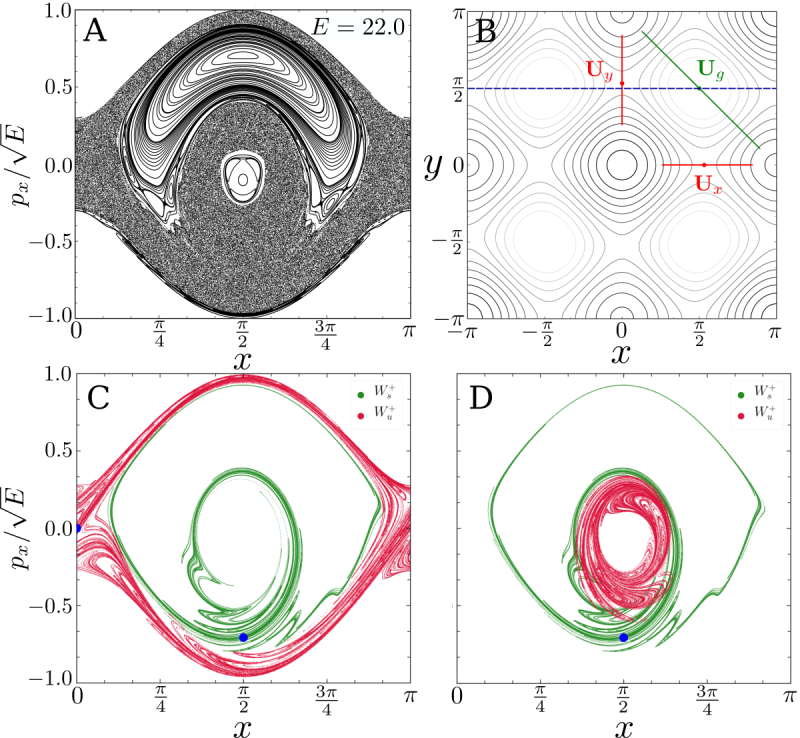}
    \caption{A) PSS $\Sigma$ for $E=22$. B) UPOs used for the calculation of the 
             manifolds displayed at frames C and D. C) Stable (green) 
             and unstable (red) manifolds from $U_g$ and $U_y$ respectively. D) Stable 
             (green) and unstable (red) manifolds from UPO $U_g$ and $U_x$ respectively. In frame 
             B, the blue dashed line depicts the spatial location of the PSS $\Sigma$ and 
             the blue dot in the remaining frames the fixed point from UPOs within it. }
    \label{fig:manifold-low-energy}
\end{figure}

As previously mentioned, the diffusion transition occurring at ($\alpha=0.1; E=36.0$) comprises a 
sudden change from ballistic ($\mu=2$) to normal ($\mu=1$) regime as seen from figure \ref{fig:diffusion_varE}, 
correlated with a simultaneous drop/growth of chaotic/regular area. At the particular energy 
$E=36.0$, the potential surface does not undergo any sudden change but its local maxima points, located 
in the unit cell at ($x=\pm\pi, y=0$) and ($x=0, y=\pm\pi$), become accessible to trajectories. 
Counterintuitively, although spatially the path for transport widens, these new equilibrium points 
act as an instability source, changing the chaotic dynamics properties by disrupting long flights 
through the lattice. It is therefore interesting to use such an orbit as reference for changes in the 
system as energy increases.

The PO along the local maxima direction exists for all energy values $E>0$, becoming 
unstable for $E \approx 6.9$ ($S_g \to U_g$) whilst branching into two other SPOs ($S_r$ and $S_y$), 
as shown in figure \ref{fig:upo-bifurcation}. This bifurcation process is the first considerable 
emergence of chaos in phase-space, as the separatrix around the new islands (related to $S_r$ and 
$S_y$) is disrupted and replaced by a chaotic layer with hyperbolic point at $U_g$ fixed point. 
In figure \ref{fig:manifold-low-energy}, portrait A shows that for energy values slightly above 
the diffusion onset ($E=22$), the chaotic layer increases and forms a connected piece, surrounding 
the three main stability islands. One may notice that the bottommost island, for 
$\frac{p_x}{\sqrt{E}} \approx 1$, with $S_r$ at its center, is highly compressed along the energy 
shell border, but still exists. 

Using the UPO $U_g$ and the ones along the saddle equilibrium points ($U_x$ and $U_y$, shown in 
portrait B of figure \ref{fig:manifold-low-energy}), their manifold structure indicates that 
for transient times the chaotic region is separately occupied by an outer layer, 
seen in portrait C as the unstable manifold of $U_y$ (in red) and an inner layer, from the stable 
manifold branch of $U_g$ (in green). 
These initially separated regions occur due to small turnstiles between the 
the manifold branches, although they appear infinitely many times for long periods, filling the whole 
chaotic sea as a single connected region. 
The unstable manifold is mostly located along regions of high $p_x$ momentum, 
related to direct flights travelling horizontally along the lattice. Due to the $x \leftrightarrow y$ 
symmetry, one can find the same division for vertical flights, from the unstable manifold of $U_x$ 
(figure \ref{fig:manifold-low-energy}, portrait D), where it is placed inside the innermost 
lobe of $U_g$ manifold, corresponding to high values of $p_y$.

\subsection{Local maxima transition}\label{sec:results-local-maxima-transition}

\begin{figure}[b]
    \centering
    \includegraphics[width=0.4\textwidth]{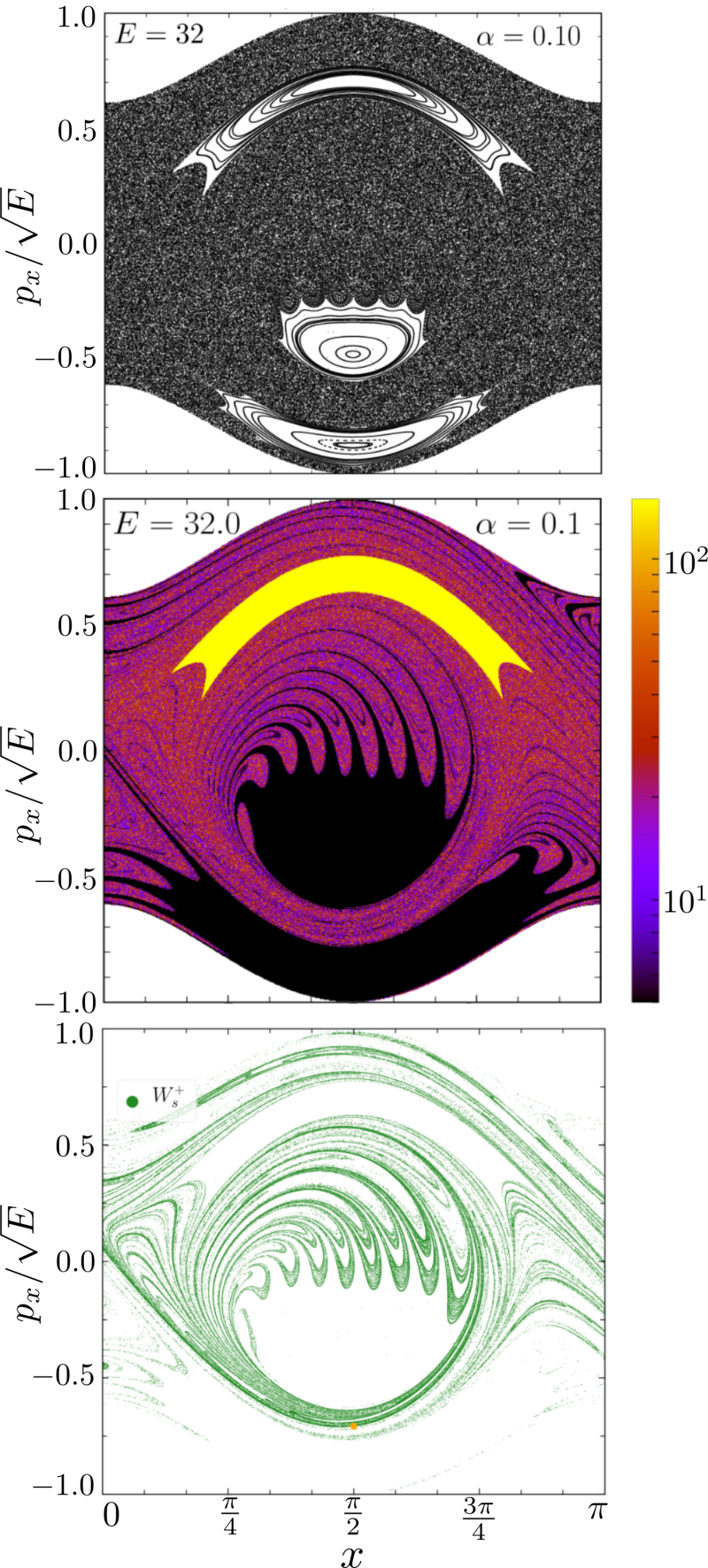}
    \caption{Phase-space portraits for $(\alpha, E)=(0.1, 32)$ with ballistic diffusion regime, 
             before transition. (Top) PSS with orbits crossing points in black. (Center) Escape 
             time basins color map. (Bottom) Stable manifolds from $U_g$.}
    \label{fig:phase-space-e32-a0.1}    
\end{figure}

\begin{figure}[b]
    \centering
    \includegraphics[width=0.4\textwidth]{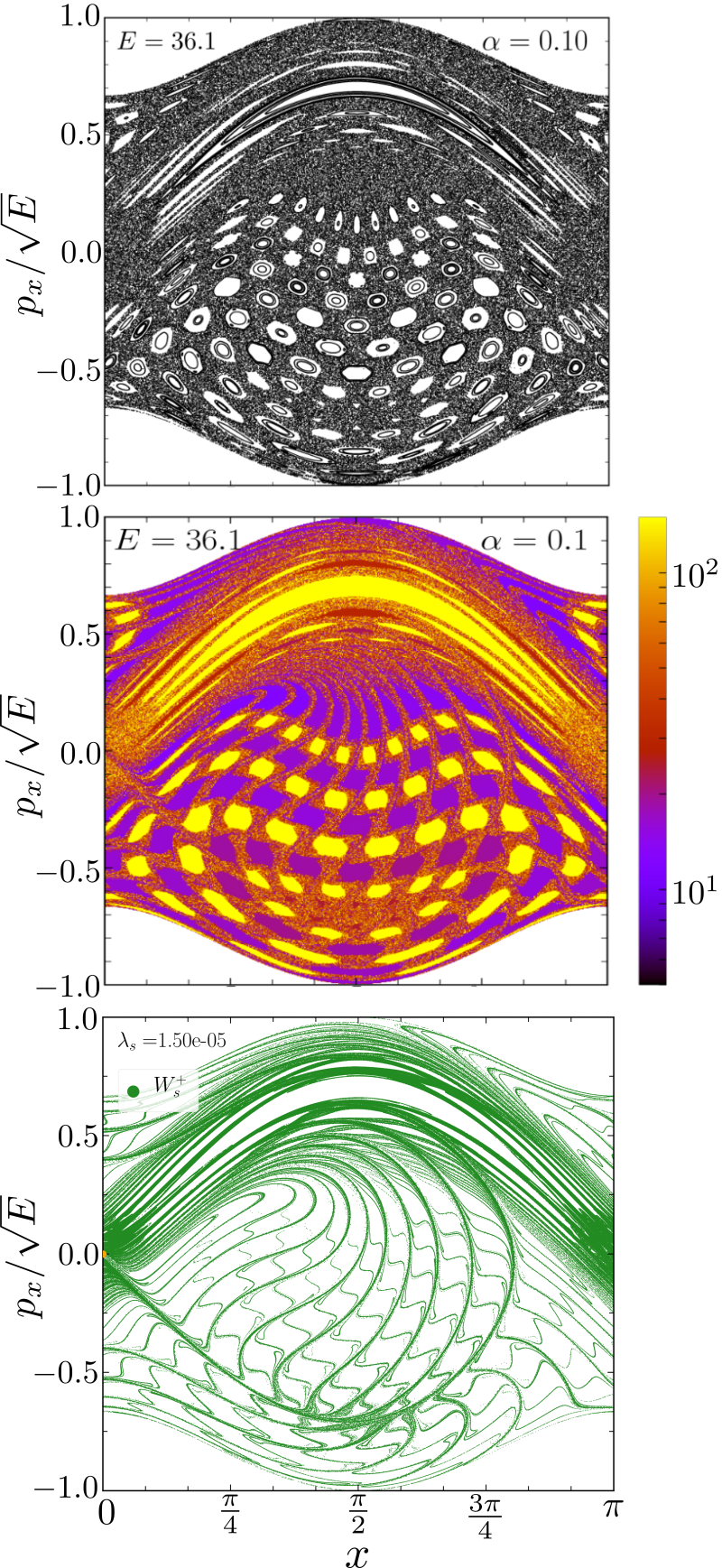}
    \caption{Same as figure \ref{fig:phase-space-e32-a0.1} but for $(\alpha, E)=(0.1, 36.1)$ energetically 
             slightly above the transition.}
    \label{fig:phase-space-e36-a0.1}
\end{figure}

\begin{figure}[b]
    \centering
    \includegraphics[width=0.4\textwidth]{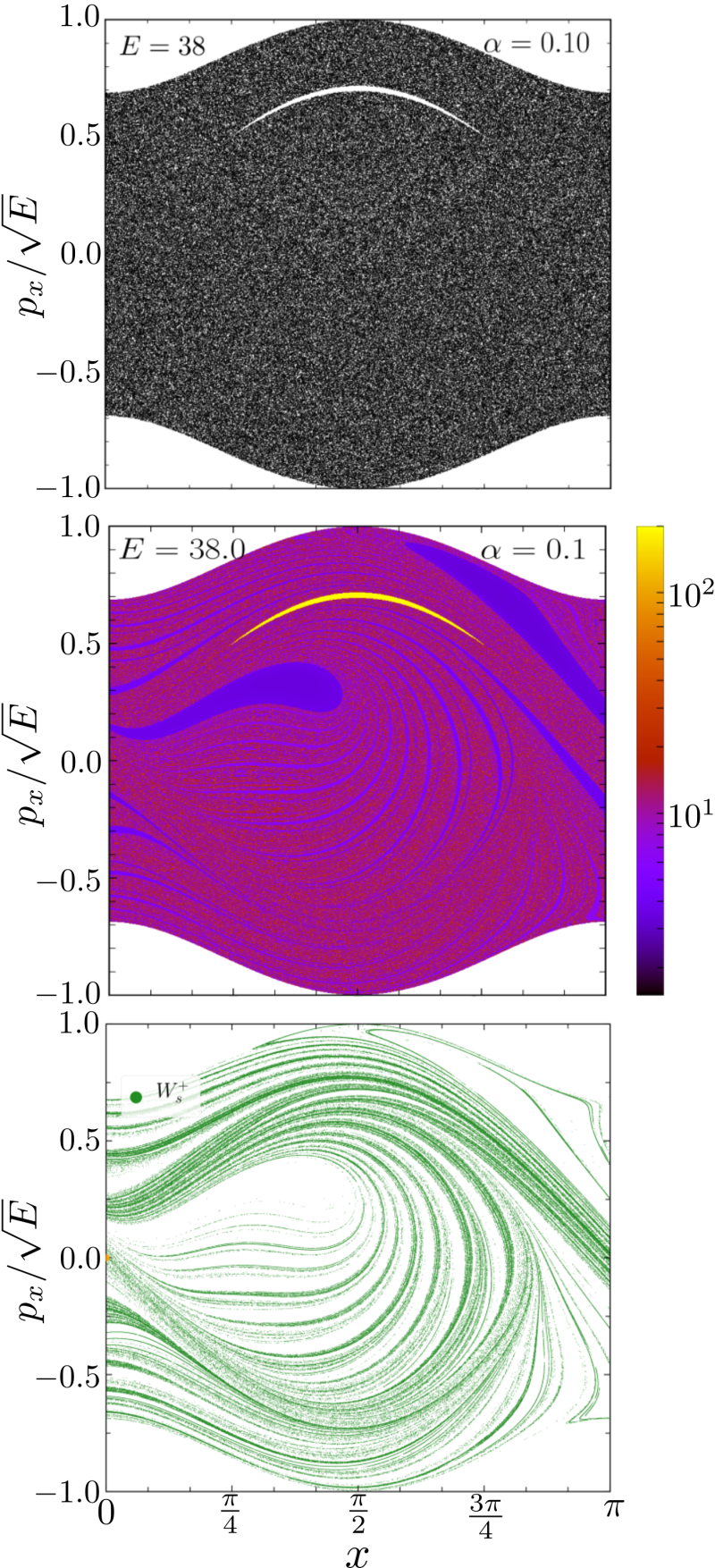}
    \caption{Same as figures \ref{fig:phase-space-e32-a0.1} and \ref{fig:phase-space-e36-a0.1} but 
             for $(\alpha, E) = (0.1, 38)$ after the transition, at normal diffusion regime.}
    \label{fig:phase-space-e38-a0.1}
\end{figure}

\begin{figure}[b]
    \includegraphics[width=0.49\textwidth]{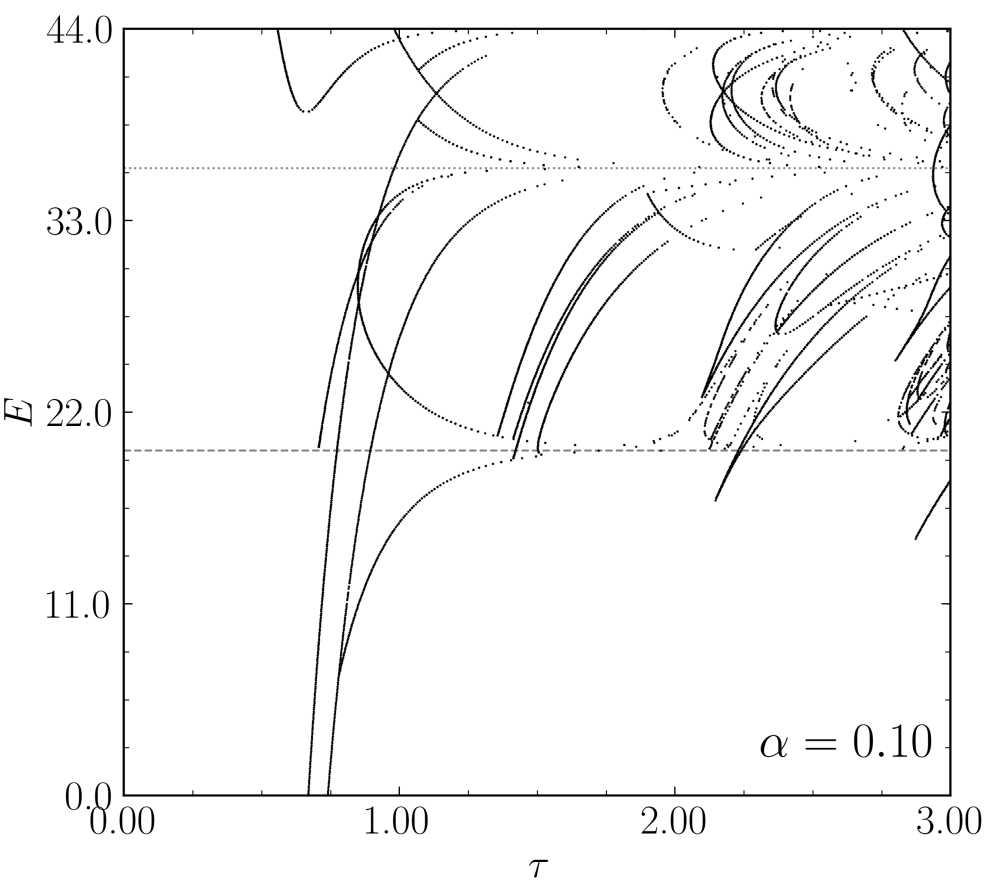}
    \caption{Period energy diagram for orbits with PSS discrete period $T=1,2$ and $3$.}
    \label{fig:period-energy-diagram}
\end{figure}

In order to detail the transition at $E=36$ and $\alpha=0.1$, phase-space portraits over the PSS $\Sigma$ 
(\ref{eq:pss}) are given for energy values before ($E=32.0$ -- Fig. \ref{fig:phase-space-e32-a0.1}), 
around ($E=36.1$ -- Fig. \ref{fig:phase-space-e36-a0.1}) and after ($E=38.0$ -- Fig. \ref{fig:phase-space-e38-a0.1}) 
the transition. These figures show the PSS with its orbit crossings (black points), the stable manifold 
from the $U_g$ UPO (Fig. \ref{fig:manifold-low-energy}), and a color map of escape time basins. In 
this context, escape time basins are simply defined as a map of the time required for initial conditions 
on the PSS $\Sigma$ to reach outside the square box with $n$ unit cells of size, \textit{i.e.}, 
$x,y \in \lt[-n\pi, n\pi\rt]$ (here $n=10$).

Before the transition, for $E=32$ (Fig. \ref{fig:phase-space-e32-a0.1}), phase-space stability regions 
are still given by the three main islands with center given by $S_y$, $S_r$ and $S_b$ SPOs (Fig. 
\ref{fig:upo-bifurcation}). As indicated by the escape time basins, the bottom islands ($S_y$ and 
$S_r$) are related to direct flights through the lattice, given their small escape time. Besides, 
channels of low escape time infiltrate the chaotic sea by stretching themselves from the instability 
region. These channels are also visible in the manifold portrait, while penetrating between the 
finger-like structure of the folded lobes that delimit the island. The upper island, related to orbits 
bounded to inside the unit cell, is the only set with ``trapped'' trajectories, whereas the 
chaotic sea alone presents an average escape time around $\langle t \rangle \approx 20$. As expected, 
the escape time basins reflect the manifolds structure, since these are the invariant lines that 
approach/depart the PSS single fixed point from the UPO $U_g$. Although not shown in figure 
\ref{fig:phase-space-e32-a0.1}, the unstable branch is perfectly symmetric, under the reflection 
$y^{\prime} = y; x^{\prime} = \pi - x$, to the stable branch. The mentioned channels may fade for higher 
box sizes or integration time, however they imply a transient behavior lasting for at least hundreds 
of time units, thus considerably long.

While the system exceeds the transition energy ($E=36$), the manifold finger-like lobes fold into 
the bottom stability islands (from $S_r$ and $S_y$), merging exactly at the fixed point position 
of the UPO $U_g$ with the stable regions vanishing in the process (figure \ref{fig:phase-space-e36-a0.1}). 
At this point, a myriad of islands emerge from the chaotic sea, forming centered chains around the unstable 
fixed point related to $U_g$, located at $(x, \frac{p_x}{\sqrt{E}})\approx(\frac{\pi}{2},-0.71)$ and also 
around the upper stability island related to $S_b$. These structures last for a narrow energy interval, 
approximately $E\in[36.0,36.3]$. 
For the group around $S_b$ island, a higher escape time (constrained orbits) can be seen, indicating a 
stickiness behavior for this region. 
The bottom chain structure, around $U_g$, presents an alternated layered structure (indicated by the 
yellow and purple islands in figure \ref{fig:phase-space-e36-a0.1}). Every layer has even discrete period, 
with each one increasing its period by 2 as they grow englobing the inner layers, forming an onion-like 
structure with the web-like manifold folding through them. 
More details on the alternating escape time range of this structure are given in appendix 
\ref{sec:appendix:po-analysis}, where the SPOs related to the chains are shown. Briefly, 
this alternation occurs due to the spatial ``closure'' of SPOs at the center of yellow 
chains, \textit{i.e.}, they return to their initial point, considering an unbounded dynamics, 
without periodic boundary conditions (PBC), thus having limited range of transport, whereas 
the chains with fast escape time (in purple) are related to SPOs that only close themselves 
when one considers spatial PBC, meaning that when unbounded, they travel as direct flights 
through the lattice. Moreover, these islands are discrete isochronous, being formed by multiple SPOs, 
with either odd or even discrete period.

After the transition, with the vanishing of the island myriad, a uniform chaotic sea mostly invades 
all phase-space, with only the upper island still remaining but in reduced size (figure \ref{fig:phase-space-e38-a0.1}). 
Amidst this chaotic region, the previous manifold web structure loses its ``horizontal'' tangling lines, 
leaving only a swirling structure, with a ``knot'' at $(x,p_x)=(0,0)$. The vanishing of the two previous 
bottom islands and their channels within the chaotic region contributes to the suppression of long flights 
and therefore transport in the system ceases from being ballistic.

\subsection{Periodic orbits bifurcations}\label{sec:results-periodic-orbits}

In order to analyse POs bifurcations and their modification in phase-space, a period-energy diagram was 
calculated with a monodromy algorithm. As developed by Baranger \textit{et al.} \cite{Baranger} and 
further detailed by Simonovi\'{c} \cite{Simonovic}, one can obtain a periodic orbit, either stable 
or unstable, from a given initial condition attempt and iteratively applying a Newton-Raphson algorithm 
to approach a periodic solution. This technique is generically applicable to any conservative Hamiltonian 
system of $N$ degrees of freedom or symplectic map. Running this procedure extensively along the PSS 
$\Sigma$, a series of POs can be found in an automated way for different energies, allowing the construction 
of a diagram displaying the dynamical period $\tau$ of the orbits found as a function of the energy value.

The result for such a diagram calculated for $\alpha=0.1$ is shown in figure \ref{fig:period-energy-diagram}, 
where POs with discrete period $T=1, 2$ and $3$ on the section $\Sigma$ are considered. The horizontal 
dashed lines mark two key energy values, namely the diffusion onset ($E=19.8$) and the potential 
local maxima ($E=36.0$). As the energy approaches either one of these values, a slowing down of the dynamics 
occurs, with the dynamical period of orbits asymptotically diverging as they reach the horizontal lines. 
This is easily understandable as the 
case of orbits reaching an unstable equilibrium point for the exact energy value of access, taking 
an infinite time span to reach it, analogous to the dynamics of a simple pendulum at the exact 
libration-rotation threshold. In this case, the saddle point between energy unit cells marks the energy 
for diffusion onset and the local maxima to the transition we studied here. Since the discrete 
periods considered in the diagram are low ($T=1,2,3$), the diagram 
does not display the island myriad appearance, given its higher period chains. However, it can be 
seen that the presence of an unstable point, even though it spreads trajectories and 
introduces chaos, is also related to a slowing down of the dynamics and the vanishing of a series of POs, 
along with the creation of completely new ones. 
Therewith, the scenario for the transition studied here is that the effect of a new instability source 
suppresses long flights within the lattice while it permeates the phase-space with chaos of more constrained 
trajectories.

\subsection{Local flights}\label{sec:results-long-transient}

\begin{figure}[b]
    \includegraphics[width=0.47\textwidth]{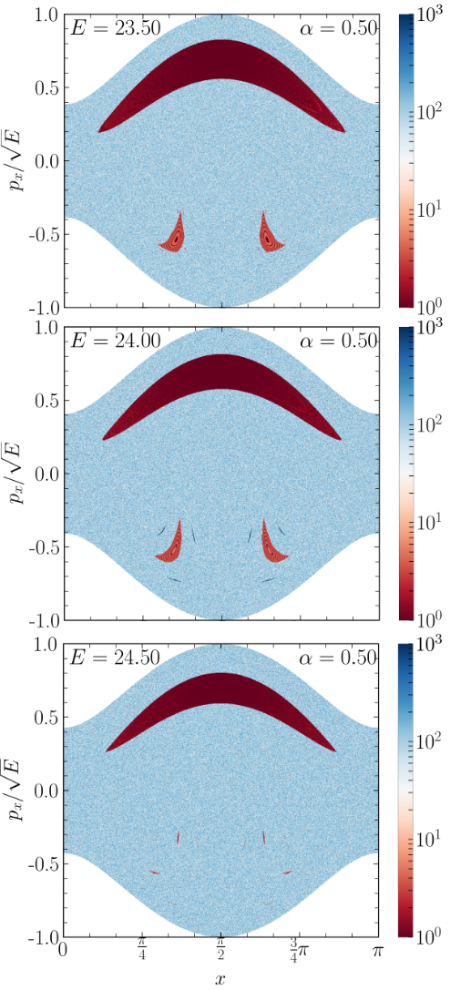}
    \caption{Displacement range color map for different energies around the sudden peak for 
             $E \approx 24$. The total integration time is $t=1000$ for each point in a 
             850$\times$830 grid.}
    \label{fig:local-flight-long-time}
\end{figure}

\begin{figure}[b]
    \centering
    \includegraphics[width=0.39\textwidth]{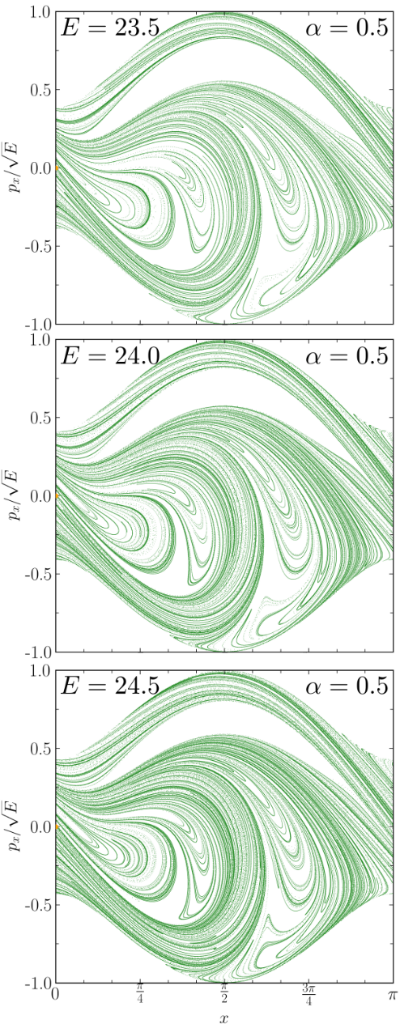}
    \caption{PSS portraits of the stable manifold branch from UPO $U_y$ for 
             different energies along the diffusion variation. Integration time 
             is $t=6.9$.}
    \label{fig:local-flight-manifold}
\end{figure}

To fully explore the aspects of the diffusion profiles obtained here (figure \ref{fig:diffusion_varE}), 
we briefly describe in this section the short peaks and valleys comprising sudden changes 
in them. These are variations in diffusion exponent uncorrelated with considerable changes in chaotic 
area, as exemplified by the energy intervals around $E \approx 24$ and $E \approx 36$ for $\alpha=0.5$, 
and $E \approx 10$, $E\approx 33$ and $E \approx 40$ for $\alpha = 1.0$.

As we did previously, we look at phase-space portraits for energies in these intervals searching for 
alterations in transport characteristics.
Therewith, to clearly display regions in phase-space with long flights, we use a displacement color map, 
similar to the escape-time ones but instead coloring each initial point with its displacement for a given 
integration time $t$ (here $t=1000$). 
Figure \ref{fig:local-flight-long-time} exemplifies this result for the transition at $E\approx 24$ and 
$\alpha=0.5$, where a peak from normal diffusion to an intermediary value ($\mu \approx 1.3$) is found. 
It can be seen that the only modification is the emergence of three small islands (shown in dark blue) 
around each of the twin islands (shown in light red) in the bottom region of the PSS, along with the 
bifurcation of these twin islands. 
The emergent island triads that appear present a very high displacement when compared to the chaotic sea 
surrounding them or the other stability islands composed of trajectories with limited range.
It is clear then that the ballistic transport due to these small islands increases the global average 
diffusion rate, making it increase quadratically but with smaller amplitude, thus taking longer 
times to increase $\mu$ but still converging towards $\mu \approx 2$.
Although not shown here, the same behavior is seen for the other peaks and variations in transport 
rate mentioned that are not correlated with changes in chaotic/regular areas.

To emphasize that the variations for the energy values mentioned above are not global transitions, 
in the sense that they are not related to major changes in the chaotic domain, 
figure \ref{fig:local-flight-manifold} shows the manifold structure for the case 
discussed above ($E\approx 24, \alpha=0.5$). 
Even though the island triads are not visible in the figure, one may notice that they appear between the 
manifold lobes without disturbing them. 
After the island triad disappears, for the highest energy value $E=24.5$, the new bifurcated islands from 
the bottommost twin islands present a small stickiness around them. Here this is indicated by the 
region unfilled by manifold lobes where these islands previously existed. However, this behavior is 
not very pronounced and is not related to the increase in displacement seen for the transport rate, given 
that the bifurcated island pairs are composed of orbits with limited range.

The effect of ballistic islands over the global diffusion rate is in agreement with the diffusion 
profiles simulated for different integration times. Whenever these islands are present, as the evolution 
time increases, the diffusion regime slowly converges towards ballistic, as shown in appendix 
\ref{sec:appendix:statistical-convergence}.
Therewith, we point out that, for the diffusion evaluation method used here, the presence of even small 
portions of phase-space with quadratic rate will imply a dominance over the total regime rate. 
Therefore the method use requires care for analysis of long times; however, it still serves as 
an indicator of the existence of long flights, given that they are indeed effects of the system dynamics. 
It is also interesting that islands with long displacement range appear amidst a 
chaotic sea with transport in a different regime, without major changes in the sea.

\section{Conclusions and perspectives}\label{sec:conclusions}

This work presented a study of classical transport of particles for a 2D lattice model based on the 
periodic potential resulting from a dipole-field interaction. It was shown that as the 
system control parameters change, namely the coupling $\alpha$ and total energy $E$, the diffusion 
exponent $\mu$, evaluated from an asymptotic law, presents sudden variations between ballistic 
and normal regimes due to the mixed nature of the system dynamics and its series of POs bifurcations.

A sudden drop of diffusion rate from ballistic ($\mu=2$) to normal ($\mu=1$) regime is found to occur 
in correlation with a decrease in chaotic area for ($\alpha=0.1, E=36.0$). A detailed 
analysis indicated that this transition occurs at an energy level for which orbits reach 
local maxima points. Even though the transport of particles seems facilitated by the wider 
spatial channel, 
the new instability source promotes a large bifurcation process with the emergence of 
multiple stable structures. Moreover, 
a global slow-down of the dynamics occurs due to orbits reaching the unstable equilibrium position, as seen in 
a period-energy diagram for 
periodic orbits.

Before the transition, long flights occur due to two main stability 
islands that vanish as $E \to 36$. At transition, phase-space becomes populated 
by an island myriad amidst a web-like manifold structure, with multiple isochronous chains 
of even discrete period. These chains form layers with increasing period and alternate 
displacement range in the lattice due to SPOs with closed or open spatial 
periodic topology.
After a narrow energy interval ($E \approx 36.3$), the island myriad vanishes and the remaining 
phase-space is dominated by a single chaotic region with long flights suppressed, presenting a normal 
diffusion regime.
This characteristic transition may occur on the system for any coupling $\alpha \neq 0$, 
although it shall be more pronounced for small values, since local maxima energy levels 
decrease as a function of $\alpha$, thus being less influent in the dynamics.

Diffusion variations that do not present correlation with chaotic area are shown to be 
caused by the emergence of small stability islands composed of long flights rather than global changes 
in dynamics. These long flights increase the average displacement towards ballistic rate but 
appearing only for long integration times, given their small weight within the orbits ensemble.

Further investigations can be made on the island myriad structure; 
it is possible that this phenomenon may occur due to a superposition of a fast dynamics, related to 
bounded orbits oscillating inside a single lattice pit, and a slow dynamics, related to the period 
divergence of UPOs reaching the unstable point \cite{Yves}. 
Moreover, different lattice topologies could be experimented and the local maxima 
transition studied for a different symmetry, such as in a honeycomb lattice, 
in order to analyse changes on the spatial closeness of the arising POs of such 
transitions.
Further investigations could include an analysis dedicated solely to the 
chaotic dynamics and its main UPOs, better detailing the influence of the unstable point 
deviation in bifurcations of POs that suppress long flights. 
Alternatively, Hamiltonian perturbations feasible within experiments, such as potential 
amplitude variation, extra monochromatic waves or noise, could be applied and compared with the 
conservative case shown here, in order to enhance the control of particles in the lattice. 

\begin{figure}[b]
    \centering
    \includegraphics[width=0.45\textwidth]{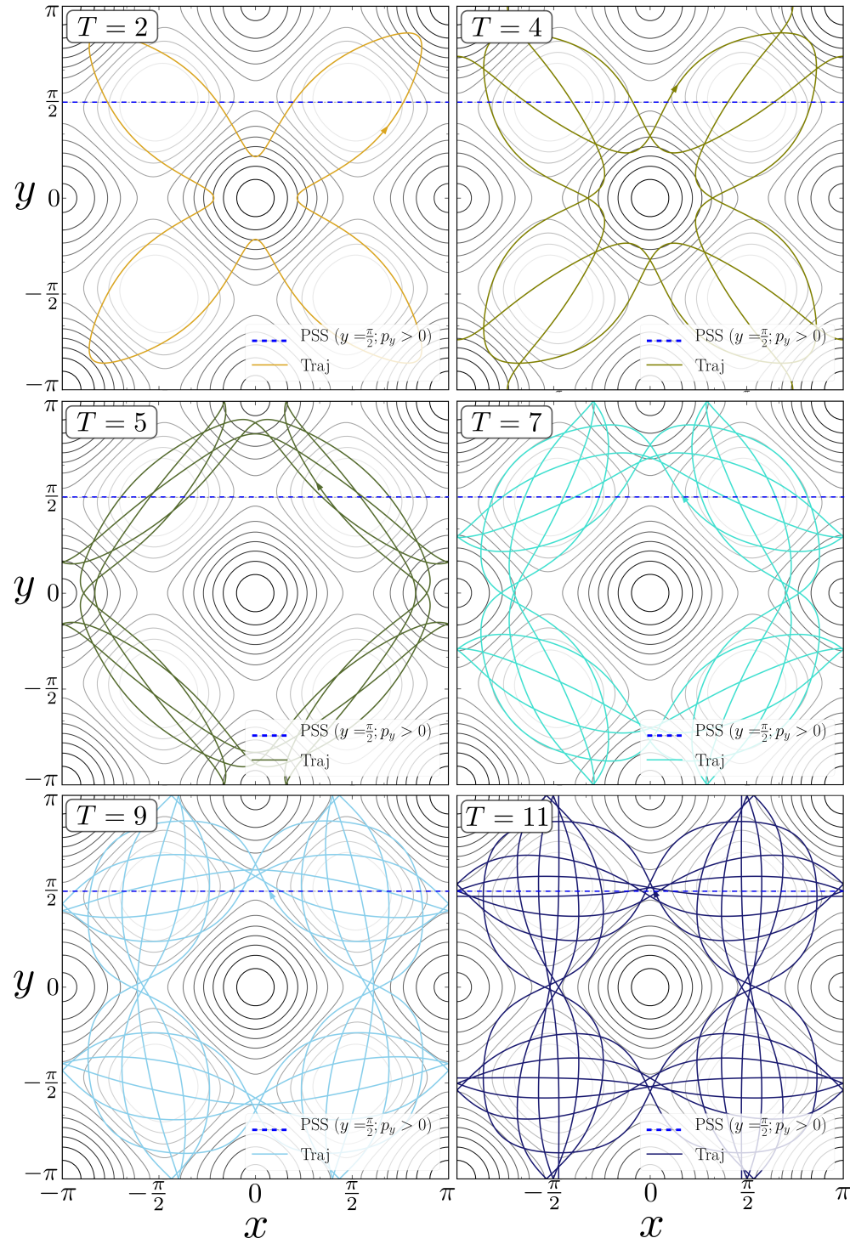}
    \caption{Selected SPOs with spatial closure, \textit{i.e.}, returning to its initial spatial 
             point without considering PBC. Orbits found for system parameters $\alpha=0.1; E=36.1$. 
             $T$ is the discrete period related to the PSS $\Sigma$.}
    \label{fig:close-spos}
\end{figure}

\begin{figure}[b]
    \centering
    \includegraphics[width=0.45\textwidth]{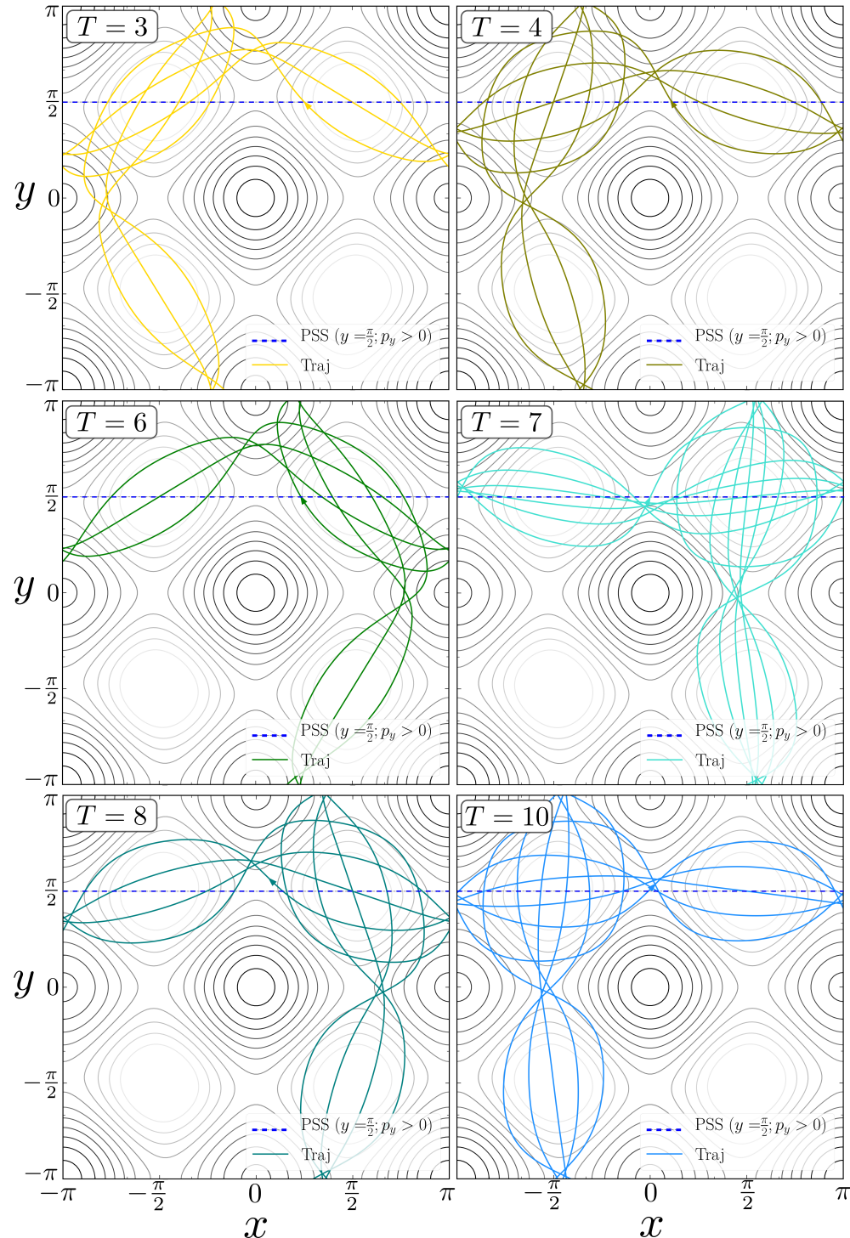}
    \caption{Selected SPOs without spatial closure, \textit{i.e.}, they only 
             return to its initial spatial point when considering PBC. Orbits found for system parameters 
             $\alpha=0.1; E=36.1$. $T$ is the discrete period related to the PSS $\Sigma$.}
    \label{fig:open-spos}
\end{figure}

\begin{acknowledgments}

M. Lazarotto would like to acknowledge Vitor M. de Oliveira for fruitful discussions and comments, 
particularly on manifold calculation and periodic orbits bifurcations.
We acknowledge the financial support from the scientific agencies:
S\~{a}o Paulo Research Foundation (FAPESP) under Grant No 2018/03211-6; 
Coordena\c{c}\~{a}o de Aperfei\c{c}oamento de Pessoal de N\'{i}vel Superior (CAPES) 
and Comit\'{e} Fran\c{c}ais d'\'{E}valuation de la Coop\'{e}ration Universitaire 
et Scientifique avec le Br\'{e}sil (COFECUB) under Grant CAPES/COFECUB 8881.143103/2017-1;
COFECUB under Grant No. 40273QA-Ph908/18;
Conselho Nacional de Desenvolvimento Cient\'{i}fico e 
Tecnol\'{o}gico (CNPq) under Grant Nos 407299/2018-1, 302665/2017-0 and 141750/2019-7.

\end{acknowledgments}

\appendix

\section*{Appendix}
This appendix presents supplementary material to the main article. It is meant to provide extra results 
to the ones shown in the main text and details on the numerical methods used for diffusion 
calculations. 
In section \ref{sec:appendix:po-analysis} we exemplify the difference between the island chains layers
that form the myriad structure by showing the different periodic orbits related to each one.
Section \ref{sec:appendix:symplectic} shows the comparison between the diffusion calculation with orbits 
integrated with a Runge-Kutta method and a symplectic method, ensuring that the average displacement 
used for the calculation of $\mu$ is accurately made.
Section \ref{sec:appendix:diffusion-var-alpha} presents extra results of diffusion profiles calculated 
for fixed energy values and $\alpha$ as variable parameter, as opposed to the cases shown in the 
main text.
Section \ref{sec:appendix:statistical-convergence} discusses the statistical convergence of the diffusion 
rate calculation and the dependence of the values of exponent $\mu$ on the simulation parameters.

\section{Periodic orbit analysis}\label{sec:appendix:po-analysis}

As shown by the escape time basins of figure \ref{fig:phase-space-e36-a0.1}, amidst the emergent 
myriad structure, island chains with even discrete period present an alternated layer structure 
of high (in yellow) and low (in purple) escape times. 
The new emergent SPOs are of two different kinds: one related to orbits with spatial closure, in the 
sense that they return to their initial position even when PBC is disabled, thus forming closed loop 
paths; 
the other group comprises orbits without spatial closure, meaning that when PBC is not considered, 
they evolve with repeating patterns without returning, thus forming 
long flights through the lattice. 
Therewith, it becomes clear 
that closed SPOs will have limited range and therefore higher escape time (or no escape at all). 
The displacement range of these orbits may increase with their discrete period but will still remain 
bounded. 
Open SPOs on the other hand will travel unboundedly through space, in an approximately ballistic way.

Figures \ref{fig:close-spos} and \ref{fig:open-spos} respectively display closed and open SPOs belonging 
to the myriad chains. It is worth mentioning that all these orbits are perfectly periodic when 
considering PBC. The periodicity of open SPOs is allowed in this case due to the periodic nature of 
the potential itself, given that a particle can return to a symmetric point in a neighbour cell, thus 
repeating the same dynamical evolution. Besides, one may notice that every orbit will have a symmetric 
counterpart, obtained by a rotations of $\frac{\pi}{2}$, with the same properties regarding stability 
(Lyapunov exponent), period and ``closedness'' but with different discrete period, given the different 
possibilities of intersection with the PSS $\Sigma$.

Moreover, the island chains are isochronous, implying that for an $N$-periodic chain, their ``links'' 
are not sequentially populated by a single SPO as usual, but instead they are composed of multiple 
SPOs with discrete period of divisors of $N$, with fixed points alternating between the links 
\cite{Sousa}. 
This may be due to the rotation symmetry of the lattice, making periodic solutions to occur in pairs 
and therefore have the same winding number. Therefore they will occur superposed in phase-space and 
form the isochronous chain structure.
As shown here, since the PSS is restricted to $x \in [0, \pi]$, some fixed points of these SPOs may 
be located in $x\in[-\pi,0]$, where the PSS map is identical to the one seen for $x>0$ but inverted, 
given the reflection translational symmetry of $\pi/2$.

\section{Symplectic integration}\label{sec:appendix:symplectic}

\begin{figure}[h]
    \centering
    \includegraphics[width=0.5\textwidth]{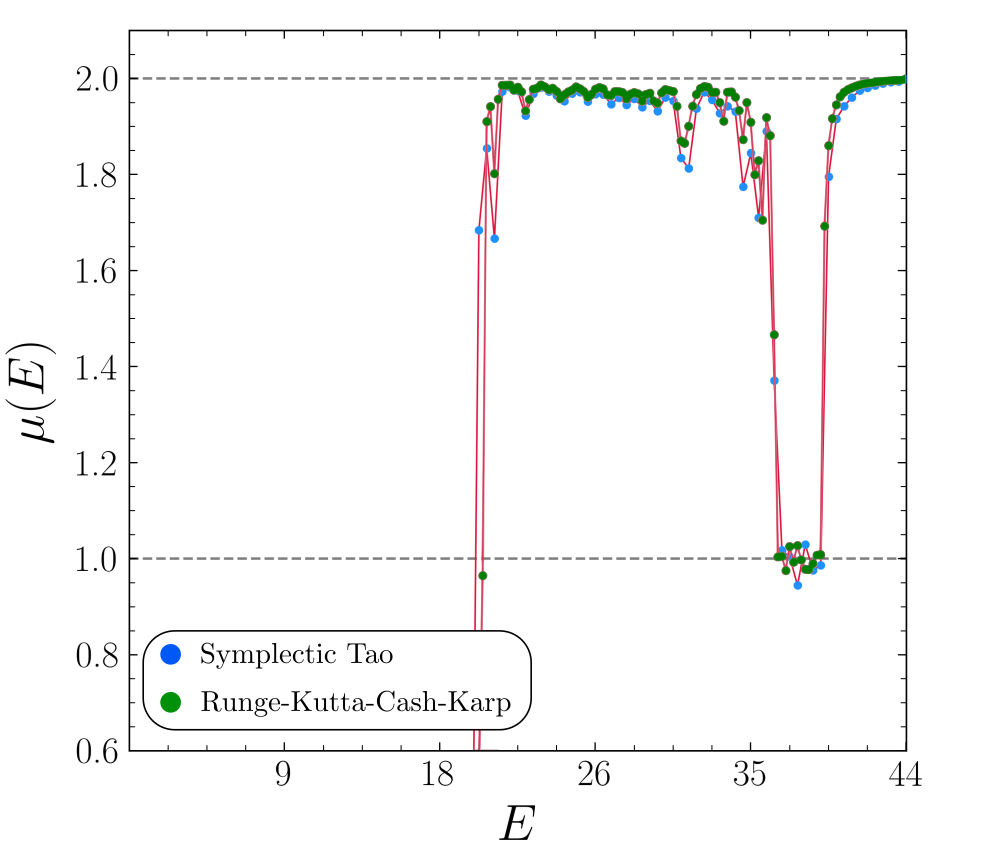}
    \caption{Comparison between diffusion exponent $\mu(E)$ calculation with RKCK method and Tao's 
             symplectic method.}
    \label{fig:diffusion_comparison_symplectic}
\end{figure}

In order to assess the quality of integration over long times required in diffusion exponent calculations, 
figure \ref{fig:diffusion_comparison_symplectic} shows the comparison between two methods: the 
Runge-Kutta-Cash-Karp (RKCK) and a symplectic one developed by M. Tao \cite{Tao}. 
The particles ensemble is randomly generated in each case, but all the remaining simulation parameters 
are equivalent, \textit{i.e.}, the total integration time $t=1000$ and ensemble size $N=10000$. The 
final values for $\mu(E)$ present a divergence of less than $0.15$ when compared between the two methods.
This indicates that even though RKCK method is not symplectic, it still yields a good average result, 
mostly due to its great energy conservation, which in this case was conserved up to order $10^{-9}$, 
regardless of deviations in particular trajectories. 
Besides, RKCK method presents faster CPU performance, which allowed the extensive calculations 
required in other analyses shown in this work.

As an adaptive time-step method, the RKCK absolute and relative precisions were selected as: 
$\epsilon_{\textrm{abs}}=\epsilon_{\textrm{rel}} = 10^{-13}$. For Tao's method, time-step is fixed 
and was selected as $dt=10^{-5}$. Also, Tao's method requires a binding factor parameter, which was set 
to $\omega=100$, since it provided the best performance regarding energy conservation (up to 
$10^{-5}$) and symplectic 2-form (up to $10^{-7}$).

\begin{figure}[h]
    \centering
    \includegraphics[width=0.44\textwidth]{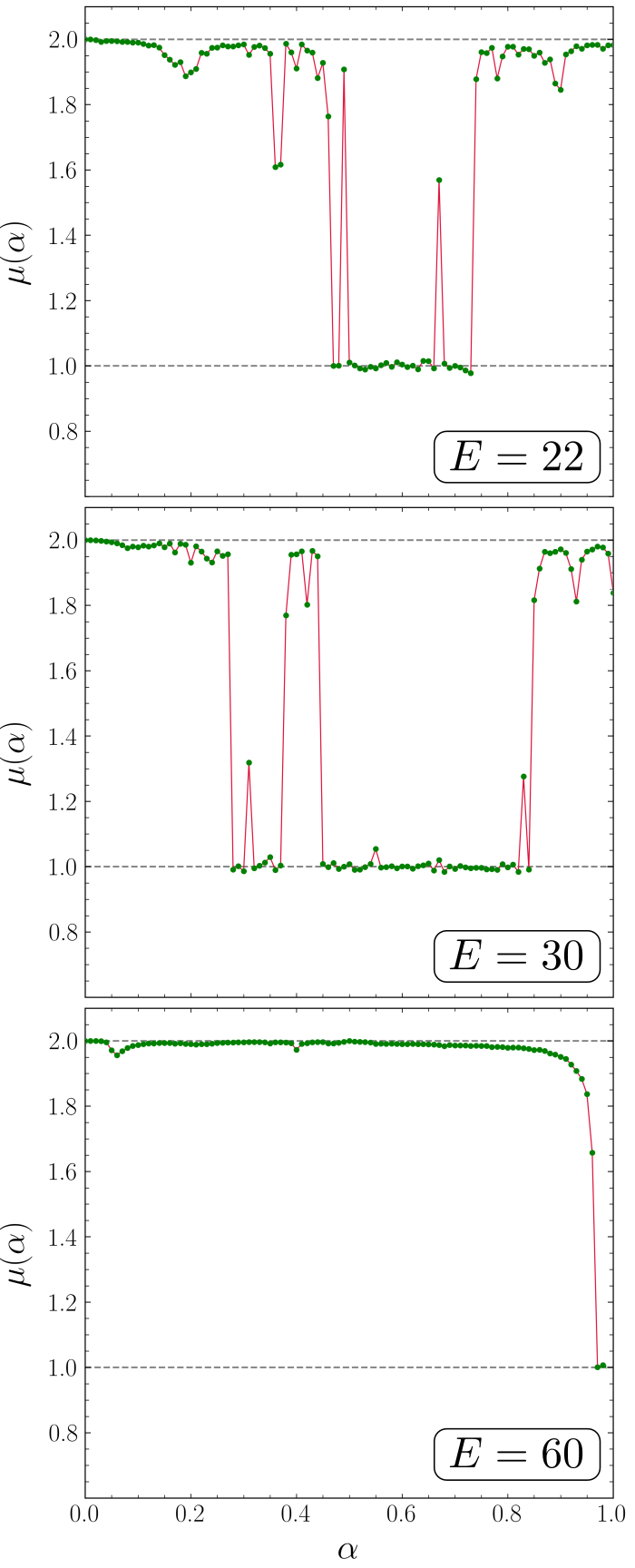}
    \caption{Diffusion exponent $\mu(\alpha)$ for different energies.}
    \label{fig:diffusion_varA}
\end{figure}

\section{Diffusion transitions}\label{sec:appendix:diffusion-var-alpha}

Figure \ref{fig:diffusion_varA} shows the calculation of diffusion exponent $\mu$ as a function 
of the coupling parameter $\alpha$ and fixed values of energy $E$, analogous to the results shown in 
the main text. The obtained profiles show the generality of the sudden transitions phenomena, 
showing that the variation of either $\alpha$ or $E$ generate the same non-trivial variations.
One may also notice that for considerably large energies (here exemplified for $E=60$), diffusion 
regime is mostly ballistic, with a sudden change to normal regime for $\alpha \approx 1.0$. 
As pointed out in the article, for high energies or small perturbations, the system behaves as two 
uncoupled pendula with rotation energy, with little impediment to diffusion. As the coupling reaches 
its maximum value, its perturbation effect becomes sufficient to influence the dynamics, since global 
maxima become energetically higher (since $V_\textrm{max} = 2U(1+\alpha)$) than the energy level of the 
particle. 
In this condition, the particle is more deflected as it moves towards maxima unstable points, instead of 
passing through them as it would if its energy was higher. These strong deflections thus may 
induce normal diffusivity through the lattice.

\section{Statistical convergence}\label{sec:appendix:statistical-convergence}

\begin{figure}[h]
    \centering
    \includegraphics[width=0.49\textwidth]{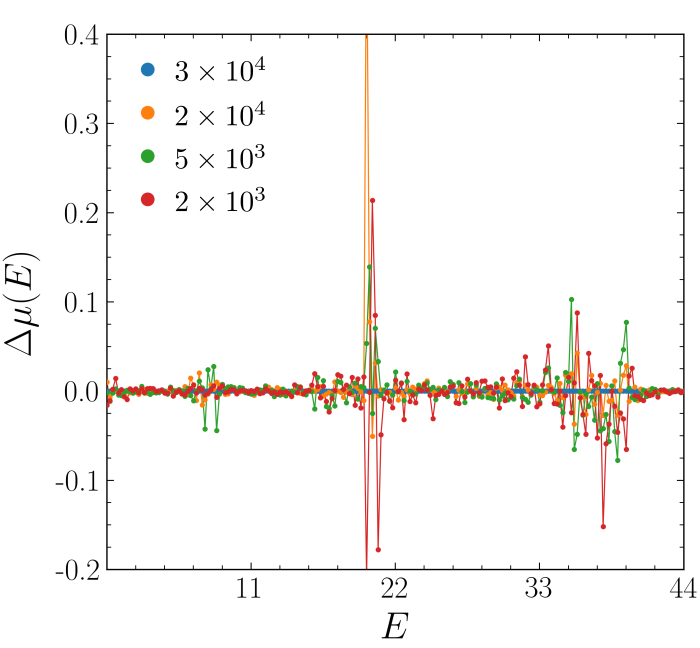}
    \caption{Diffusion exponent difference $\Delta\mu=\mu_N - \mu_n$ between different values of ensemble size 
             $n$ and the best case for $N=3\times10^4$. In this case, $t=3000$ and $\alpha=0.1$.}
    \label{fig:diffusion-stat-ensemble-size}
\end{figure}

\begin{figure}[h]
    \includegraphics[width=0.41\textwidth]{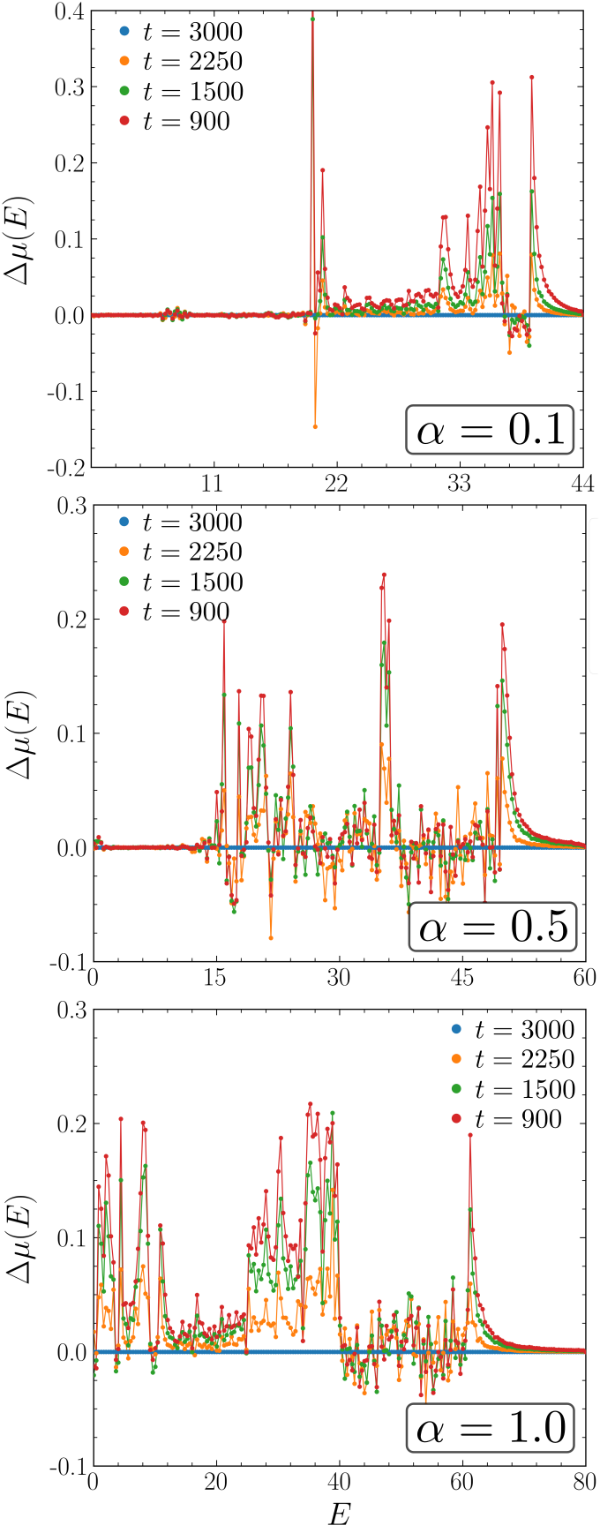}
    \caption{Diffusion exponent difference $\Delta\mu = \mu_T - \mu_t$ between different values of integration time 
             $t$ and the best case for $T=3000$. In all cases, $N=3 \times 10^4$.}
    \label{fig:diffusion-stat-integration-time}
\end{figure}

In order to ensure the statistical reliability of the calculated diffusion profiles, this section 
discusses the dependence of the values of $\mu$ with the total integration time $t$ and the ensemble 
size $N$. 
This is done by comparing the simulation in the best case scenario (highest $t$ or $N$) with multiple 
other ones with lower values of $t$ and $N$. As mentioned, the best case scenario here corresponds to 
the parameters used in the results presented in the main article, namely $t=3\times10^3$ and $N=3\times10^4$.
It can be asserted then that the fluctuations and transitions found in the 
diffusion rate as a function of either $\alpha$ or $E$ are not due to statistical imprecision, but 
indeed caused by changes in the dynamics. It can also indicate that the values selected for $t$ 
and $N$ ensure good convergence of the final result.

Figure \ref{fig:diffusion-stat-ensemble-size} shows the plot for the difference $\Delta\mu = \mu_N - \mu_n$, 
with $\mu_N$ evaluated for the highest ensemble size and $\mu_n$ for a smaller one, for different $n$.
In all cases: $\alpha=0.1$ and $t=3\times10^3$.
One can notice that the amplitudes of $\Delta\mu$ for different $n$ did not show to be large, with 
the average difference being of order $\Delta\mu\approx0.05$. A particularly higher deviation 
occurs around the energy value for diffusion onset, corresponding to $E=19.8$ for $\alpha=0.1$, although 
this is expected given the absolute amplitude variation of $\mu$ itself. 
Even though the test presented here comprises only a single value of $\alpha$, it can be 
expected that for other values this result still holds. One may also notice that simulations 
could be optimized by selecting $N$ at least one order of magnitude smaller than the value used 
in the main article, with the same qualitative final result.

Similarly, figure \ref{fig:diffusion-stat-integration-time} shows the same procedure but for 
different integration times. Here the comparison is made as $\Delta\mu = \mu_T - \mu_t$, with $\mu_T$ 
obtained for the longest integration and $\mu_t$ for any smaller value. 
The results indicate a higher amplitude variation for the different $t$, but still below 0.15 in average. 
As for the case of different $N$, the same sensitivity is shown for energy values around transitions 
or high fluctuations of $\mu$.
Nonetheless, the data for different integration times present a trend of increasing values of $\mu(E)$ 
towards ballistic regime as the integration time increases. This is due to the presence of particles 
moving at ballistic rate that dominate the average displacement of the ensemble.
However, the same result points out that convergence is close and the results found here are valid, 
given the small amplitude for $\Delta\mu$.

In general, regardless of the sensitivity of $\Delta\mu$ around regions with high variation
of $\mu$, one can see that they occur at the same energy values. 
This allows us to assert that the transitions found in the exponent $\mu$ as we change the 
system parameters are not mere statistical variation artifacts. Also, some fluctuations become 
apparent just after a long transient; in fact they are not transitions properly said, in the sense 
that topological changes occur in phase-space, but just a long time transient behavior due the 
appearance of small islands composed of long flights that end up dominating the ensemble average 
displacement.

\bibliographystyle{unsrt}
\bibliography{ref.bib}

\end{document}